\documentclass{article}
\usepackage{amsfonts}
\usepackage{amssymb}
\usepackage{amsmath}
\usepackage{graphicx}
\usepackage{epstopdf}
\usepackage{slashed}
\usepackage{setspace}

\newcommand{\reef}[1]{(\ref{#1})}
\newcommand{\rom}[1]{\mathrm{#1}}

\newcommand{\cd}{{\cal D}}

\newcommand{\cn}{{\cal N}}

\newcommand{\cf}{{\cal F}}

\newcommand{\co}{{\cal O}}

\newcommand{\be}{\begin{equation}}
\newcommand{\ee}{\end{equation}}

\newcommand{\Khoze}[1]{W_{#1}}

\def\be{\begin{equation}}
\def\ee{\end{equation}}
\def\bea{\begin{eqnarray}}
\def\eea{\end{eqnarray}}
\def\ba{\begin{array}}
\def\ea{\end{array}}
\def\bd{\begin{displaymath}}
\def\ed{\end{displaymath}}

\def\a{\alpha}
\def\b{\beta}

\def\d{\delta}
\def\e{\epsilon}

\def\g{\gamma}
\def\h{\eta}

\def\m{\mu}

\def\th{\theta}

\def\pa{\partial}

\def\>{\rangle} 
\def\<{\langle} 
\def\Dsl{D \hskip-.6em \raise1pt\hbox{$ / $ } }
\def\to{\rightarrow}

\def\pa{\partial}

\def\lab{\label}

\newcommand{\eps}{\epsilon}
\newcommand{\lra}{\leftrightarrow}

\def\da{{\dot\alpha}}

\def\tQ{\tilde{Q}}

\begin{document}

\setstretch{1.05}

\begin{titlepage}

\begin{flushright}
MIT-CTP-3971 \\
IPMU 08-0054 \\
\end{flushright}
\vspace{1cm}

\begin{center}
{\Large\bf Recursion Relations, Generating Functions,} \\[2mm]
{\Large\bf and Unitarity Sums in $\cn =4$ SYM Theory}  \\
\vspace{1cm}
{\bf Henriette Elvang${}^{a}$,
Daniel Z.~Freedman${}^{a,b}$, Michael Kiermaier$^{a,c}$} \\
\vspace{0.7cm}
{${}^{a}${\it Center for Theoretical Physics}\\
{${}^{b}${\it Department of Mathematics}}\\
         {\it Massachusetts Institute of Technology}\\
         {\it 77 Massachusetts Avenue}\\
         {\it Cambridge, MA 02139, USA}}\\[5mm]
{\it {${}^c$ Institute for the Physics and Mathematics of the Universe}}\\
{\it {University of Tokyo}}\\
{\it {Kashiwa, Chiba 277-8582, Japan}}\\[5mm]
{\small \tt  elvang@lns.mit.edu,
 dzf@math.mit.edu, mkiermai@mit.edu}
\end{center}
\vskip .3truecm

\begin{abstract}
We prove that the MHV vertex expansion is valid for any NMHV tree amplitude of $\cn = 4$ SYM. The proof uses induction to show that there always exists a complex deformation of three external momenta such that the amplitude falls off at least as fast as $1/z$ for large $z$. This validates the generating function for $n$-point NMHV tree amplitudes. We also develop generating functions for anti-MHV and anti-NMHV amplitudes.
As an application, we use these generating functions to evaluate several examples of intermediate state sums on unitarity cuts of 1-, 2-, 3- and 4-loop amplitudes.
In a separate analysis, we extend the recent results of arXiv:0808.0504 to prove that there exists a valid 2-line shift for any $n$-point tree amplitude of $\cn = 4$ SYM.
This implies that there is a BCFW recursion relation for any tree amplitude of the theory.

\end{abstract}

\end{titlepage}

\setstretch{0.5}
\tableofcontents
\setstretch{1.05}

\newpage

\setcounter{equation}{0}
\section{Introduction}

Recursion relations for tree amplitudes based on the original
 constructions of CSW \cite{csw} and BCFW \cite{bcf,bcfw}  have had
 many applications to QCD,  $\cn =4$ SYM theory,  general
 relativity, and  $\cn =8$ supergravity.\footnote{Readers are referred to the review \cite{bdkreview}  and the references listed there.}
In this paper we are concerned with recursion relations for
$n$-point tree amplitudes $A_n(1,2,\dots,n)$ in which the
external particles can be any set of gluons, gluinos, and
scalars of $\cn =4$ SYM theory.

Recursion relations follow from the analyticity and
pole factorization of tree amplitudes in a complex variable $z$
associated with a deformation or shift of the external momenta.
A valid recursion relation requires that the shifted amplitude vanishes as
$z \to \infty.$ This has been proven for external gluons by
several interesting techniques, \cite{csw,bcfw,nima1}, but  there is only partial information for
amplitudes involving other types of particles \cite{bdp}.  Of particular relevance to us is a very recent result of Cheung
 \cite{ccg} who shows that there always exists at least one  valid 2-line shift
 for any amplitude of the $\cn = 4$  theory in which one particle is a negative helicity gluon and the other $n-1$ particles are arbitrary. We use SUSY Ward identities to
extend the result to include amplitudes with $n$ particles of any\footnote{With the exception of one 4-scalar amplitude. See Sec.~\ref{s:2line}.} type.
 Thus for $\cn=4$ SYM amplitudes there always exists a valid 2-line shift
which leads to a recursion relation of the BCFW type. It is simplest for MHV amplitudes but provides a correct representation of all amplitudes.

For NMHV amplitudes the MHV vertex expansion of   CSW is usually preferred, and this is our main focus. The MHV vertex expansion is associated with a 3-line shift \cite{risager}, and it is again required to show that amplitudes vanish at large $z$ under such a shift. To prove this,
we use the BCFW representation to study $n$-point NMHV amplitudes in which one particle is a negative helicity gluon but other particles are arbitrary. Using induction on $n$ we show that there is always at least one 3-line shift for which these amplitudes vanish in the large $z$ limit.  The restriction that one particle is a negative helicity gluon can then be removed  using SUSY Ward identities.  Thus there is a valid (and unique, as we argue) MHV vertex expansion for any NMHV amplitude of the $\cn =4$ theory.

Next we turn our attention to the generating functions which have been devised
to determine the dependence of amplitudes on  external states of the theory.
The original and simplest case is the MHV generating function of Nair \cite{nair}.
A very useful extension to diagrams of the CSW expansion for NMHV
amplitudes was proposed by Georgiou, Glover, and Khoze \cite{ggk}.   MHV and NMHV generating functions were further studied in a recent paper \cite{BEF} involving two of the present authors.  A 1:1 correspondence was established between the particles of $\cn =4$ SYM theory and differential operators involving the Grassmann  variables of the generating function.  MHV
amplitudes are obtained  by applying products of these differential operators of
total order 8 to the generating function, and  an NMHV amplitude is obtained by applying a product of operators of total order 12.  We review these constructions  and emphasize that the NMHV
generating function has the  property that every 12th order differential operator
projects out the correct  MHV vertex expansion of the corresponding  amplitude, specifically the expansion which is validated by the study of the large $z$ behavior of 3-line shifts described above.
In this sense the NMHV generating function is universal in $\cn =4$ SYM theory. Its form does not contain any reference  to a shift,  but  every amplitude is produced in the expansion which was established using a valid 3-line shift.

In \cite{BEF} it was  shown in examples at the MHV level how the generating function formalism automates and simplifies the sum over intermediate helicity states required to compute the unitarity cuts of loop diagrams.  In this paper we show how Grassmann integration  further simplifies and extends the MHV level helicity sums.  We then apply the universal generating function to examples of helicity sums involving MHV and NMHV amplitudes.

Even
 in the computation of MHV amplitudes 
at low loop order, N$^2$MHV and N$^3$MHV
 tree 
amplitudes are sometimes required to complete the sums over intermediate states.
 We derive generating functions for all N$^k$MHV amplitudes in \cite{efk2}  (see also \cite{Drummond:2008cr}).
Note though that when the amplitudes have $k+4$ external lines these are equivalent to anti-MHV or
anti-NMHV amplitudes. In this paper we discuss
a general procedure to
convert the conjugate of any N$^k$MHV generating function into an
anti-N$^k$MHV generating function which can be used to compute spin
sums. We study the $n$-point anti-MHV and anti-NMHV cases in detail and apply
them in several examples of helicity sums. These include 3-loop and 4-loop cases.

We use conventions and notation given in Appendix A of \cite{BEF}.


 \setcounter{equation}{0}
 \section{$\cn = 4$ SUSY Ward identities}

The bosons and fermions of $\cn = 4$ SYM  theory can be described by
the following annihilation operators, which are listed in order of
descending helicity:
\be \lab{n4ops}
 B_+(i)\, , \hspace{5mm}
 F^a_+(i)\, , \hspace{5mm}
 B^{ab}(i) = {1\over 2}\, \,\e^{abcd}B_{cd}(i)\, ,\hspace{5mm}
 F_a^-(i)\, ,\hspace{5mm}
 B^-(i) \, .
\ee
The argument $i$ is shorthand for the 4-momentum $p^\m_i$ carried by the
particle.
Particles of opposite helicity transform in conjugate representations of the
$SU(4)$ global symmetry group (with indices $a,b,\dots$), and scalars satisfy
the indicated $SU(4)$ self-duality condition.
In this paper it is convenient to ``dualize" the lower
 indices of positive  helicity annihilators and introduce a notation in which
 all particles carry upper $SU(4)$ indices, namely:
  \bea \lab{newops}
 &&
 A(i)\,=\, B_+(i) \, , \hspace{7mm}
 A^a(i)\,=\,F^a_+(i)\,  , \hspace{7mm}
 A^{ab}(i) \,=\, B^{ab}(i)\,  ,\\[2mm]  \nonumber
&& \hspace{4mm}
 A^{abc}(i)\,=\,\e^{abcd} F_d^-(i) \,  ,\hspace{9mm}
 A^{abcd}(i) \,=\, \e^{abcd} B^-(i)   \,.
\eea
Note that the helicity (hence bose-fermi statistics) of any particle is then determined by the $SU(4)$ tensor rank $r$ of the operator $A^{a_1\dots a_r}(i)$\,.


Chiral supercharges
 $Q^a \equiv - \e^\a\,Q^a_\a$ and
$\tilde{Q}_a \equiv \tilde{\e}_{\dot{\a}} \tilde{Q}^{\dot{\a}}_a$ are defined to
include contraction with the  anti-commuting parameters $\e^\a, \tilde{\e}_{\dot{\a}}$ of SUSY transformations.
The commutators of the operators $Q^a$
and $\tilde{Q}_a$ with the various annihilators are given by:
\bea
\begin{array}{rcl}
\big[\tQ_a,A(i)\big] &=& 0\, ,\\[2mm]
\big[\tQ_a,A^b(i)\big] &=& \<\epsilon\, i\>\,\d^b_a\,A(i)\, ,\\[2mm]
\big[\tQ_a,A^{bc}(p)\big]
&=&\<\epsilon\,i\> \, 2! \, \d^{[b}_a\,A^{\raisebox{0.6mm}{\scriptsize$c]$}}(i)\, ,\\[2mm]
\big[\tQ_a,A^{bcd}(i)\big] &=&\<\epsilon\, i\>\, 3!\, \d_a^{[b} A^{\raisebox{0.6mm}{\scriptsize$cd]$}}(i)\, ,\\[2mm]
\big[\tQ_a,A^{bcde}(i)\big] &=& \<\epsilon\,i\>\, 4!\, \d_a^{[b} A^{\raisebox{0.6mm}{\scriptsize$cde]$}}(i) \, ,
\end{array}
\hspace{8mm}
\begin{array}{rcl}
[Q^a,A(i)] &=& [i\,\epsilon]\,A^a(i)\, , \\[2mm]
\big[Q^a,A^b(i)\big] &=& [i\,\epsilon]\, A^{ab}(i)\, , \\[2mm]
\big[Q^a,A^{bc}(i)\big]
&=&[i\,\epsilon]\, A^{abc}(i)\, , \\[2mm]
\big[Q^a,A^{bcd}(i)\big]&=& [i\,\epsilon]\,A^{abcd}(i)\, , \\[2mm]
\big[Q^a,A^{bcde}(i)\big] &=& 0\, .
\end{array}
\lab{n4tQ}
\eea
Note that $\tQ_a$  raises the helicity of all operators and involves
the spinor angle
 brackets 
$\<\e\,i\>$.  Similarly, $Q^a$
lowers the helicity and spinor square brackets $[i\, \e]$ appear.

It is frequently useful to suppress indices and simply use $\co(i)$ for any
annihilation operator from the set in \reef{newops}.  A generic $n$-point amplitude may then be denoted by
\be \lab{amp1}
A_n(1,2,\dots,n) = \big\<\co(1) \co(2)\ldots\co(n)\big\>\,.
\ee
$SU(4)$ invariance  requires that the total number of (suppressed) indices
is a multiple of 4, i.e.   $\sum_{i=1}^n r_i\,=\,4m$.

It is well known, however, that
amplitudes $A_n$ with $n \ge 4$ vanish if $\sum_{i=1}^n r_i\,=\,4$.  To see this we use  SUSY Ward identities,  as in the particular case
\bea
 \nonumber
0 &=&
\big\<\big[\tQ_1\,,\, A^1(1) A^{1234}(2) A(3)\ldots  A(n)\big]\big\>
\\[1mm]
 &=& \<\e\, 1\>\big\<A(1) A^{1234}(2) A(3)\ldots A(n)\big\>\,
 +\,\<\e\, 2\> \big\< A^1(1) A^{234}(2) A(3)\ldots A(n)
 \big\> \, .  
\lab{wi1}
\eea
There are exactly two terms in the Ward identity.  One can choose $|\e\> \sim |2\>$ and learn that the first amplitude, involving one negative helicity  and $n-1$ positive
helicity gluons, vanishes.  The second fermion pair amplitude must then also vanish.\footnote{This argument  does not apply to the case $n=3$  because  the strong constraint of momentum conservation forces either $|1\> = c |2\>$ or $|1] =c  |2]$.  If the first occurs, then one cannot  choose $|\e\>$ so as to isolate one
of the two terms in \reef{wi1}, and $A_3(1,2,3)$ with $\sum_{i=1}^n r_i\,=\,4$ need not vanish for complex momenta. This amplitude is anti-MHV.}
We have chosen one specific example for clarity, but the argument applies to
all amplitudes with  $\sum_{i=1}^n r_i\,=\,4$ and $n \ge 4$.  To see this
consider
\be \lab{wi2}
0\,=\, \big\<\big[\tQ_1\,,\, \co(1) \co(2) \ldots  \co(n)\big]\big\>\,.
\ee
$SU(4)$ symmetry requires that the upper index  1 appears exactly twice among the operators $\co(i)$ and that the indices 2,3,4 each
 appear 
once.
The commutator again contains two terms, one from each $\co(i)$ that carries the index 1.  The argument above then applies immediately.

Let's continue and discuss the Ward identity \reef{wi2} for the general case
$\sum_{i=1}^n r_i\,=\,4m,~ m \ge 2$.  The upper index 1 must appear $m+1$ times among the $\co(i)$ and the indices 2,3,4 each appear $m$ times. The
commutator then contains $m+1$ terms,  and each of these involves an amplitude  with $\sum_{i=1}^n r_i\,=\,4m$.

Ward identities with the conjugate supercharges $Q^a$ have the similar structure
\be \lab{wi3}
0\,=\, \big\<\big[Q^1\,,\, \co(1) \co(2) \ldots  \co(n)\big]\big\>\,.
\ee
This is a non-trivial identity if the index 1 appears $m-1$ times, and the indices
2,3,4 each appear $m$ times. The commutator then contains $n-m+1$ terms,
each again with an amplitude with   $\sum_{i=1}^n r_i\,=\,4m$.
To summarize, all amplitudes related by any one SUSY Ward identity must have the same total number of upper $SU(4)$
indices.  It is then easy to see the case $m=2$ corresponds to MHV amplitudes,
$m=3$ to NMHV,  while general N$^k$MHV amplitudes must carry a total of
$4(k+2)$ upper indices.

In \cite{BEF}
a 1:1 correspondence
between annihilation operators  in \reef{newops} and differential operators involving the Grassmann variables $\eta_{ia}$ of generating functions
 was introduced.
We will need this correspondence in Sec.~\ref{s:gf} below, so we restate it here:
\bea\lab{opcorr}
&&A(i) \lra 1\, ,~~~~~~~
A^a(i)  \lra \frac{\pa}{\pa\h_{ia}}\, ,
~~~~~~~A^{ab}(i)\lra \frac{\pa^2}
{\pa\h_{ia}\pa\h_{ib}}\, ,\\
&& A^{abc}(i)\lra \frac{\pa^3}{\pa\h_{ia}\pa\h_{ib}\pa\h_{ic}}\, ,~~~~~~~
A^{abcd}(i)\lra \frac{\pa^4}{\pa\h_{ia}\pa\h_{ib}\pa\h_{ic}\pa\h_{id}}\,.
\nonumber
\eea
Thus a particle state whose upper $SU(4)$ rank is $r$ corresponds to a differential operator of order $r$.  In accord with \cite{BEF} we
will refer to the rank $r$ as the $\h$-count of the particle state. We showed in \cite{BEF} that an MHV amplitude containing a given set of external particles can be obtained by applying a product of the corresponding differential operators of total order 8
to the MHV generating function, and NMHV amplitudes that are obtained by
applying products of total order 12 to the NMHV generating function.  The classification of amplitudes based on the total $\h$-count of the particles they contain is a consequence of $SU(4)$ invariance.

\setcounter{equation}{0}
\section{Valid 3-line shifts  for NMHV amplitudes}
\lab{s:valid3}

The major goal of this section is to prove that
there is at least one 3-line shift for \emph{any} NMHV amplitude  $A_n(m_1,\ldots,m_2,\ldots, m_3,\ldots)$ under which the amplitude vanishes at the rate $1/z$ or faster as $z \to\infty$. We show that this is true when
the shifted lines $m_1,\,m_2,\,m_3$ share at least one common $SU(4)$ index, and that such a shift is always available.  The first step in the proof is to show that there is a valid 3-line shift for any NMHV amplitude $A_n(1^-,\ldots,m_2,\ldots, m_3,\ldots,n)$,
with particle 1 a negative helicity gluon, $m_2$ and $m_3$ sharing a common $SU(4)$ index, and the other states arbitrary.
This requires an intricate inductive argument which we outline here and explain in further detail in Appendix~\ref{app17}.
We then generalize
the result to arbitrary NMHV amplitudes using a rather short argument based on SUSY Ward identities.
This result implies that there is a valid  MHV vertex expansion for any NMHV amplitude in $\cn = 4$ SYM theory.

\subsection{Valid shifts for $A_n(1^-,\ldots,m_2,\ldots, m_3,\ldots,n)$}

We  must start with a correct representation of the amplitude $A_n(1^-,\ldots,m_2,\ldots, m_3,\ldots,n)$ which we can use to study the limit $z \to\infty$
under the
3-line shift
\cite{risager}
of the spinors
 $|1],\,|m_2],\, |m_3]$  
given by
\bea
\nonumber
|1] &\to& ~\,|\hat{1}]  ~=~ |1]\,+\, z\, \<m_2 m_3\> \, |X]\, ,\\
\lab{3shiftTX}
|m_2]&\to& |\hat{m}_2] = |m_2]\,+ \,z\, \<m_3\,1\>\, |X]\, ,\\
\nonumber
|m_3]&\to& |\hat{m}_3]  =|m_3]\,+ \,z\, \<1\,m_2\>\, |X]\,,
\eea
where $|X]$ is an arbitrary reference spinor.  Angle bracket
spinors $|1\>,\, |m_2\>,\,|m_3\>$ are not shifted.
It is assumed that the states $m_2$ and $m_3$ share at least one common $SU(4)$ index.
We must show that the large $z$ limit of the amplitude deformed by this shift vanishes for all $|X]$. The amplitude then contains no pole at $\infty$ and Cauchy's theorem can be applied to derive  a recursion relation containing a sum of diagrams, each of which is a product of
two MHV subdiagrams connected by one internal line.  This recursion relation
agrees with the MHV vertex expansion of \cite{csw}.

The representation we need was recently established by Cheung \cite{ccg}
who showed that every amplitude $A_n(1^-,\dots,x,\dots,n)$, with particle 1 a negative helicity gluon and others arbitrary, vanishes in the large $z$ limit of the 2-line shift
\be\lab{2shift}
  |\tilde{1}] = |1] + z |x] \, ,~~~~ |\tilde{1}\> = |1\> \, , ~~~~~~~~~~~
  |\tilde{x}] = |x] \, ,~~~~|\tilde{x}\> = |x\> - z |1\> \,.
\ee
This leads to a  recursion relation containing a sum of diagrams which are each products of a Left subdiagram, whose $n_L$ lines include the shifted line $\tilde{1}$ and a Right subdiagram whose $n_R$ lines include $\tilde{x} $.
Clearly, $n_L +n_R = n+2$. See Fig.~\ref{fig:2shift}.
\begin{figure}
\begin{center}
 \includegraphics[width=6.5cm]{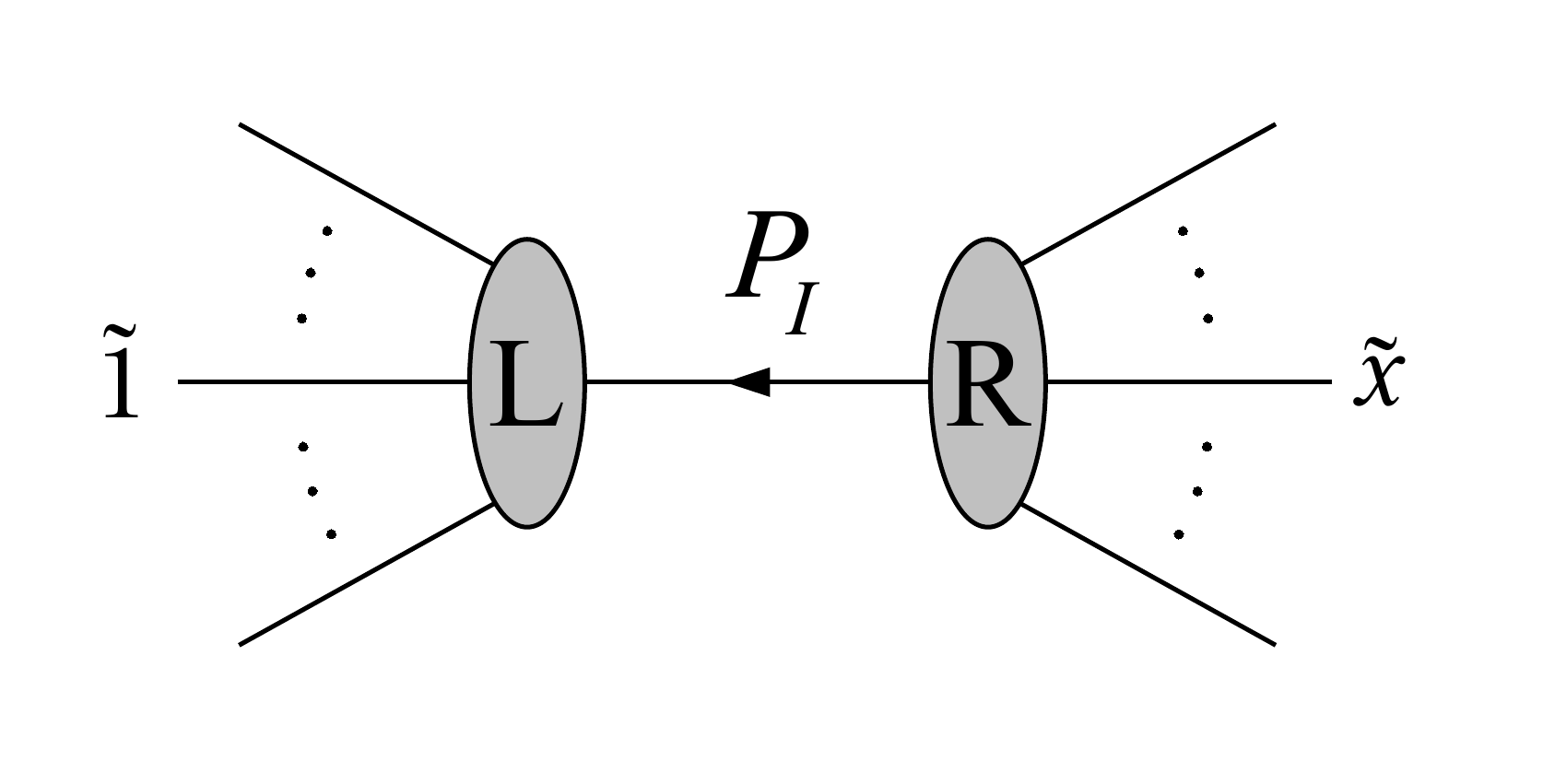}\\[3mm]
\end{center}
\vspace{-8mm}
\caption{
Diagrammatic expansion of an amplitude $A_n(1^-,\dots,x,\dots,n)$ under a 2-line shift $[1^-,x\>$.}
\lab{fig:2shift}
\end{figure}
As explained in Appendix~\ref{app17},
\emph{only the following two types of diagrams contribute} to the recursion relation:
\begin{description}
\item[Type A:] MHV $\times$ MHV diagrams with $n_L \ge 3$ and $n_R \ge 4$.
\item[Type B:] NMHV $\times$ anti-MHV diagrams with $n_L =n-1$ and $n_R =3$.
\end{description}
Our strategy is to consider the effect of the shift \reef{3shiftTX} as a secondary shift on each diagram of the recursion relation above.  The action of the shift depends on how $m_2$ and $m_3$ are placed on the left and right subdiagrams.
In Appendix~\ref{app17}, we show that every Type A diagram  vanishes as $z \to \infty$,
and that Type B diagrams can be controlled by induction on $n$.
Thus the full amplitude $A_n(1^-,\ldots,m_2,\ldots, m_3,\ldots,n)$,
 with lines $m_2$ and $m_3$ sharing a common SU(4) index, 
 is shown to fall off at least as fast as $1/z$ under the
 $[1,m_2,m_3|$-shift.
The
full argument is complex and requires detailed examination of  special cases for
$n=6,7$.  Interested readers are referred to Appendix~\ref{app17}.

We used a 2-line shift simply to have a correct representation of the amplitude to work with
in the proof of the large $z$ falloff. That shift plays no further role. In the following we use a more general designation in which line 1 is relabeled $m_1$.


\subsection{Valid shifts for $A_n(m_1,\ldots,m_2,\ldots, m_3,\ldots)$}

We now wish to show that any NMHV amplitude vanishes at least as fast as
$1/z$ under the 3-line shift
\be\label{3lineshift}
  |\hat{m}_1] = |m_1] + z \<m_2 m_3 \> |X]\, ,\qquad
  |\hat{m}_2] = |m_2] + z \<m_3 m_1\> |X]\, ,\qquad
  |\hat{m}_3] = |m_3] + z \<m_1 m_2 \> |X]\, ,
\ee
provided that the 3 lines $m_1,m_2,m_3$ have at least one common $SU(4)$ index which we denote by $a$. In the previous section we showed that the shift is valid if $r_1 = 4$,
 where, as usual, $r_1$ denotes the $\eta$-count of line $m_1$.

We work with
SUSY Ward identities and proceed by (finite, downward) induction on $r_1$.
We assume that $1/z$ falloff holds for all amplitudes with
$r_1=\bar r$,
 for some $1\leq\bar r\leq 4$.
We now want to show that it also holds for amplitudes with $r_1=\bar r-1$.  Since  $r_1 <4$, there is at least one  $SU(4)$ index not carried by the annihilation operator $\co(m_1)$. We denote this index by $b$ and use $\co^b(m_1)$ to denote the operator of rank $\bar r$ containing the original indices of $\co(m_1)$ plus $b$.
 This operator satisfies
 $[\tQ_b,\co^b(m_1)]=\<\e\,m_1\> \co(m_1)$ (no sum on $b$). The Ward identity  we need (with $|\e\>$  chosen such that $\<\e\,m_1\>\ne 0$)
is
\bea
\nonumber
  0&=&\<[\tQ_b\,,\,  \co^b(m_1)\dots\co(m_2)\ldots\co(m_3)\dots]\>\\[1mm]
  &=&
  \<\e\, m_1\>\<\co(m_1)\dots\co(m_2)\ldots\co(m_3)\dots\> \,+
  \,\<\co^b(m_1)[\tQ_b,\dots\co(m_2)\ldots\co(m_3)\dots]\>\,.
\lab{wi4}
\eea
The first term in the final equality contains the NMHV amplitude we are
interested in (which is an amplitude with $r_1 = \bar r - 1$).  The index $b$ appears 3 times among the operators $\co(i)$ in the commutator in the second term, so there are 3
 potentially
non-vanishing terms in that commutator.  Each term contains an unshifted angle bracket $\<\e \,i\>$ times an NMHV amplitude with
 $r_1=\bar r$ and lines $m_1,m_2,m_3$ sharing the common index $a$, thus
it is an amplitude which vanishes as $1/z$ or faster.  We conclude
\begin{equation}
    \boxed{
    \text{
    \begin{tabular}{c}
    $A_n=
    \big\<\co(m_1)\dots\co(m_2)\ldots\co(m_3)\dots\big\>
    \to\frac{1}{z}$  ~~under the 3-line shift \reef{3lineshift} \\[2mm]
    if lines $m_1$, $m_2$, $m_3$ share at least one $SU(4)$ index\,.~~
    \end{tabular}
    }}
\end{equation}

We have
 thus
established valid $[m_1,m_2,m_3|$-shifts if the common index criterion is satisfied.
One may ask if this is a necessary condition.
There are examples of shifts not satisfying our criterion but which still
produce $1/z$ falloff. In  \cite{BEF} the falloff of the 6-gluon amplitude $A_6(1^-,2^-,3^-,4^+,5^+,6^+)$ under 3-line shifts was studied
numerically.
The results in (6.49) of \cite{BEF}
 show 
that some
shifts of three lines which do not share a common index do nonetheless give $1/z$ falloff while others are $O(1)$ at large $z$.

Note that  the 6-gluon amplitude above has a unique shift satisfying our criterion, while any 6-point NMHV amplitude in which the 12 indices appear on 4 or more lines has several such shifts.
The case  $A_6 = \<A^{1234}(1) A^{1234}(2)A^{123}(3)A^4(4)A(5)A(6)\>$ is
one of many examples. Both $[1,2,3|$ and $[1,2,4|$ are valid shifts
 in this case.


\setcounter{equation}{0}
\section{Generating functions}
\label{s:gf}

A generating function for
 MHV tree amplitudes 
in $\cn =4$ SYM theory was invented  by Nair \cite{nair}. The construction was extended to the
NMHV level by Georgiu, Glover, and Khoze \cite{ggk}. Generating functions are a very convenient way
to encode how an amplitude depends on the helicity and
global symmetry
 charges
of the
external states. The generating function
for an $n$-point amplitude depends on $4n$
real Grassmann variables
 $\h_{ia}$, and the spinors $|i\>$, $|i]$ and momenta $p_i$
of the external lines.   A 1:1 correspondence
between states of the theory and Grassmann derivatives was defined in \cite{BEF} and given above in \reef{opcorr}.  Any desired amplitude is obtained by
applying the product of
 differential
operators associated with its external particles to the
generating function. It was also shown in \cite{BEF} that amplitudes obtained from the generating function obey SUSY Ward identities.

The discussion below is in part a review, but we emphasize the shift-independent universal property of the NMHV generating function.
We follow \cite{BEF}, and more information can be found in that reference.

\subsection{MHV generating function}

The MHV generating function is\footnote{
 Lines are identified  periodically, $i \equiv i+n$.}
 \be
\lab{n4gen} F_n =\Big(\prod_{i=1}^{n}
 \<i,i+1\> 
\Big)^{-1}\d^{(8)}\Big(\sum_{i=1}^{n}|i\>\eta_{ia}\Big)\,.
\ee
The 8-dimensional $\d$-function can be expressed as the product of its arguments, i.e.
\be\lab{del8}
 \d^{(8)}\Big(\sum_{i=1}^{n}|i\>\eta_{ia}\Big)\,=\,
 \frac{1}{16}\prod_{a=1}^{4} \sum_{i,j=1}^{n}\<i\,j\>\,\eta_{ia}\,\eta_{ja}\,.
\ee

In Sec.~2 we saw that MHV amplitudes $\<\co(1)\co(2)\ldots\co(n)\>$  contain products
of operators with a total of 8 $SU(4)$ indices (with each index value appearing
exactly twice among the $\co(i)$).  The associated product of differential
operators $D_i$ from \reef{opcorr} has total order 8 and the amplitude may be expressed as:
\bea  \lab{mhvamp}
\<\co(1)\co(2)\ldots\co(n)\>&=&  D_1D_2\ldots D_n \,F_n\\
&=&  \frac{\<~\>\<~\>\<~\>\<~\>}{\<12\>\<23\>\ldots
 \<n-1,n\> 
\<n 1\>}
\eea
The numerator is the spin factor which is the product of 4 angle brackets from
the differentiation of $\d^{(8)}$.  It is easy \cite{BEF} to compute spin factors.
Here is an example of a 5-point function:
\be \lab{exmhv}
\<A^{1234}(1)A^1(2)A^{23}(3)A(4)A^4(5)\>\,=\,\frac{\<12\>\<13\>^2\<15\>}{\<12\>\<23\>\<34\>\<45\>\<51\>}
\ee
Like brackets are not cancelled because we want to illustrate how this example conforms to the general structure above.

\subsection{The MHV vertex expansion of an NMHV amplitude }

The NMHV generating function is closely tied to the MHV vertex expansion of \cite{csw}.
The diagrams of such an expansion contain  products of two MHV subamplitudes with at least one shifted line in each factor.  For  $n$-gluon NMHV amplitudes it was shown in
\cite{risager} that this expansion agrees with the recursion relation obtained
from the 3-line shift \reef{3lineshift}.  For a general NMHV amplitude the
recursion relation from any valid shift also leads to an expansion containing diagrams with two shifted MHV subamplitudes.
In $\cn =4$ SYM theory this expansion has the following important property
which we demonstrate below;  \emph{the recursion relation obtained from any valid 3-line shift contains no reference to the shift used to derive it.}
Therefore, all $[m_1,m_2,m_3|$-shifts in which the shifted lines contain at least one common $SU(4)$ index yield the same recursion relation!
 The MHV vertex expansion is thus unique for every amplitude.

A typical MHV vertex diagram is illustrated in
Fig.~\ref{fig:3shift}.
\begin{figure}
\begin{center}
 \includegraphics[width=6.5cm]{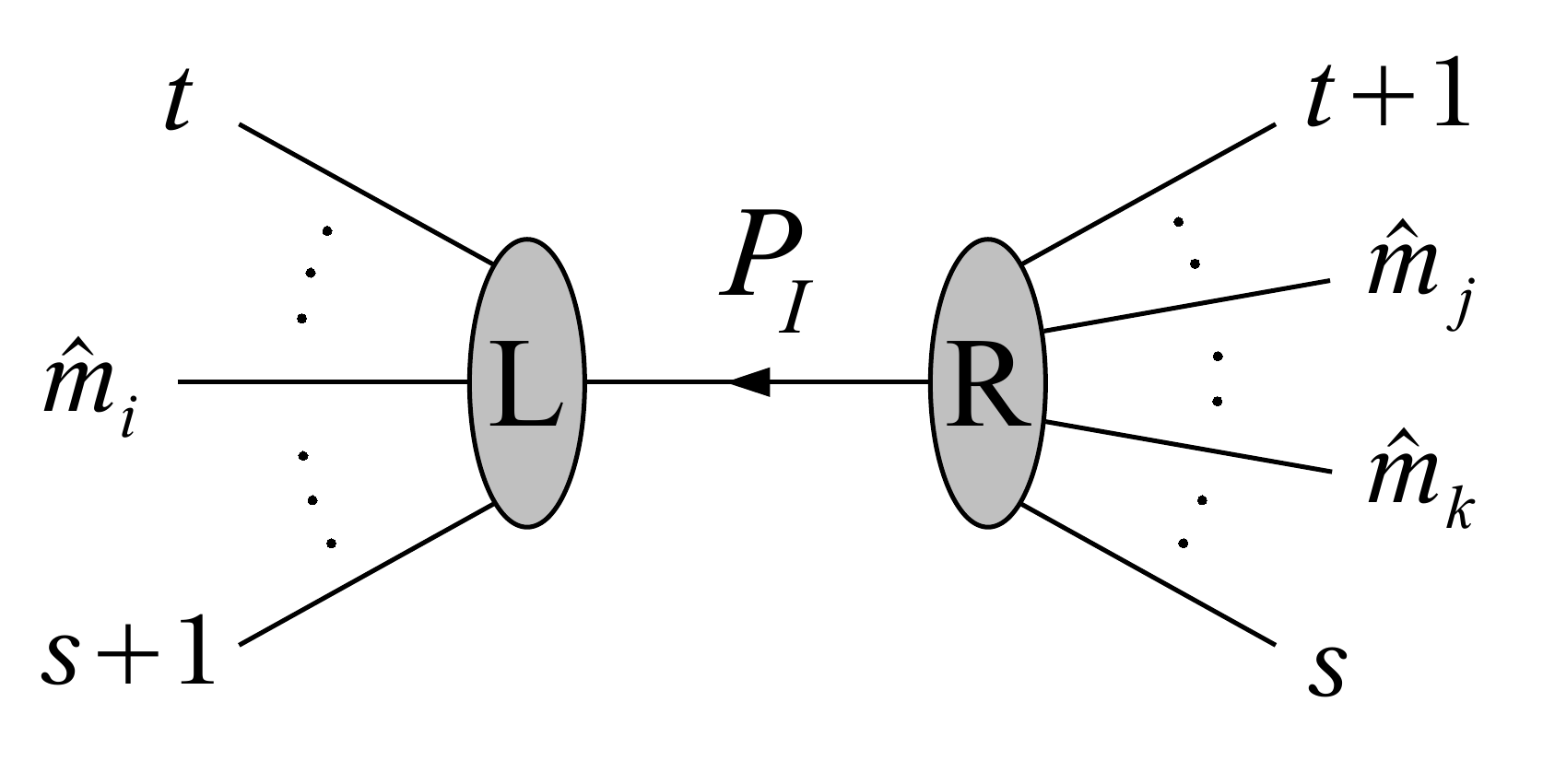}\\[3mm]
\end{center}
\vspace{-6mm}
\caption{A generic MHV vertex diagram of an NMHV amplitude $A_n(m_1,\dots,m_2,\dots,m_3,\dots)$, arising from a 3-line shift $[m_1,m_2,m_3|$. The set of lines $\hat{m}_i$, $\hat{m}_j$, $\hat{m}_k$ is a cyclic permutation of $\hat{m}_1$, $\hat{m}_2$, $\hat{m}_3$.}
\lab{fig:3shift}
\end{figure}
In our conventions all particle lines are regarded as outgoing.
Therefore, if the particle on the internal line
 carries a particular set of
$SU(4)$ indices of rank $r_I$ in the left subamplitude,  it must carry the complementary set
of indices of rank $4-r_I$ in the right amplitude. Since each subamplitude must be
$SU(4)$ invariant, there is a unique state of the theory which can propagate across the
diagram. Any common index $a$ of the shifted lines $m_1,m_2,m_3$, must also appear on the internal particle in the subdiagram that contains only one shifted line.

Let us assume, as
indicated in Fig.~\ref{fig:3shift}, that the left subamplitude contains the external lines
$s+1,\dots,t$, including a shifted line $\hat{m}_i$, and that the right subamplitude contains lines $t+1,\ldots,s$ and the remaining shifted lines $\hat{m}_j, \hat{m}_k$ (here $i,j,k$ denotes a cyclic permutation of $1,2,3$).
In each subamplitude one uses
the CSW prescription for the
angle spinor of the internal line:
\be \lab{cswp}
|P_I \> \equiv P_I |X] = \sum_{i=s+1}^t |i\>[i\,X]\,, \qquad  |-P_I\>= - |P_I \>\,.
\ee

The contribution of the diagram to the  expansion is simply the product
of the MHV subamplitudes times the  propagator  of the
internal line.  It is given by:
\be \lab{mhvdi}
\frac{(\<~\>\<~\>\<~\>\<~\>)_L}
{\<-P_I,s+1\>\cdots
 \<t-1,t\>
\<t,-P_I\>}\frac{1}{P_I^2}\frac{(\<~\>\<~\>\<~\>\<~\>)_R}{\<P_I,t+1\>\cdots\<s-1,s\>\<s, P_I\>}
\ee
The numerator factors are products of 4 angle brackets which are the
spin factors for the left and right subamplitudes. They depend on the spinors $|i\>$ and
$|\pm P_I\>$ in each subamplitude and can be calculated easily from the MHV generating function   described in Sec. 4.1.
The denominators contain the
same cyclic products of $\<i ,i+1\>$ well known from the Parke-Taylor formula \cite{pt}, and the
standard propagator factor $ P_I^2 = (p_{s+1}+\ldots+p_t)^2$.

The main point is that there is simply no trace of the initial shift in the entire
formula \reef{mhvdi} because
\begin{itemize}
\item[i.] only angle brackets are involved, and they are
unshifted, and
\item[ii.] the propagator factor is unshifted.
\end{itemize}

To complete the
discussion we suppose that there is a another valid shift on lines
 $[m_1',m_2',m_3'|$
 which have a common index we will call $b$. Consider any diagram that appears in the
expansion arising from the original
 $[m_1,m_2,m_3|$-shift. If each subdiagram happens to
contain (at least) one of the $m_i'$ lines, then the same diagram with the same
contribution to the amplitude occurs in the expansion obtained from the $m_i'$ shift. A
diagram from
 the
$m_i$ expansion in which all 3 $m_i'$ lines are located
 in one of the two
subamplitudes cannot occur because the index $b$ would appear 3 times in that
subamplitude. This is impossible because that subamplitude is MHV and contains each
$SU(4)$ index only twice. This completes the argument that  any valid 3-line
shift yields the same MHV vertex expansion in which the contribution of each
diagram is independent of the chosen shift.
The MHV vertex expansion of any NMHV amplitude is unique.

The contribution of each diagram to the expansion depends on the
reference spinor $|X]$.  Since the physical amplitude contains no such
arbitrary object, the sum of all diagrams must be independent of $|X]$.  This
important fact is guaranteed by the derivation of the recursion relation provided that the amplitude vanishes as $z \to \infty$ for all $|X]$. This is what we proved in section \ref{s:valid3}.

\subsection{The universal NMHV generating function}

To obtain the generating function for the (typical)
MHV vertex diagram in Fig.~\ref{fig:3shift} we start with the product of MHV generating functions for each sub-diagram times the internal propagator. We rewrite this
product as
\be \lab{nmhvprod}
 \frac{1}{ \prod_{i=1}^{n} \<i, i+1\>} \, W_I \, \d^{(8)}(L)\d^{(8)}(R)
\ee
with
\bea  \lab{details}
W_I &=&   \frac{\<s,s+1\>\<t,t+1\> }{\<s\,P_I\>\<s+1,P_I\>\,P_I^2\,\<t\,P_I\>\<t+1,P_I\>}\label{Khoze}\\
L &=&  |-P_I\>\,\h_{Ia}+ \sum_{i=s+1}^{t} |i\>\,\h_{ia} \\
R &=&  |P_I\>\,\h_{Ia} +  \sum_{j=t+1}^{s} |j\>\,\h_{ja} \,.
\eea
The Grassmann variable $\h_{Ia}$ is used for the internal line.
We have separated the denominator factors in \reef{mhvdi} into a Parke-Taylor cyclic product over the full set of external lines times a factor
$W_I$ involving  the left-right split, as used in \cite{ggk}.

The
contribution of
\reef{nmhvprod}
to the diagram for a given process is then obtained
by applying
the appropriate product of Grassmann derivatives from \reef{opcorr}. This product includes derivatives for external lines and the derivatives $D_{I_L} D_{I_R}$ for the internal lines.  It follows from the discussion above that
the operators $D_{I_L}$ and $D_{I_R}$   are of order   $r_I$ and
$4-r_I$ respectively, and that their product is simply
\be  \lab{prod}
D_{I_L} D_{I_R}\,=\, \prod_{a=1}^{4} \frac{\pa}{\pa\h_{Ia}}\,.
\ee
We apply this 4th order derivative to \reef{nmhvprod}, convert the derivative to
a Grassmann integral as in \cite{BEF},
and integrate using the formula \cite{ggk}
\be  \lab{etaint}
  \int \prod_{a=1}^4 d\h_{Ia}\,
  \d^{(8)}\big( L \big) \,
  \d^{(8)}\big( R \big)
  ~=~
  \d^{(8)}\Big(\sum_{i=1}^{n}|i\>\h_{ia}\Big)
  \prod_{b=1}^4\sum_{j=s+1}^{t}\<P_I\,j\>\h_{jb} \, .
\ee
Thus we obtain the generating function
\bea
\nonumber
\cf_{I,n} &=&\frac{ \d^{(8)}\Big(\sum_{i=1}^{n}|i\>\h_{ia}\Big)}{ \prod_{i=1}^{n} \<i, i+1\>}~W_I  \prod_{b=1}^4\sum_{i=s+1}^{t}\<P_I\,i\>\h_{ib} \\
 &=&\frac{ \d^{(8)}\Big(\sum_{i=1}^{n}|i\>\h_{ia}\Big)}{ \prod_{i=1}^{n} \<i, i+1\>}~W_I  \prod_{b=1}^4\sum_{j=t+1}^{s}\<P_I\,j\>\h_{jb}
 \lab{nmhvgen}
\eea
The two expressions are equal because  $\d^{(8)}$ for the external lines
is present.  Using \reef{del8} one can see that \reef{nmhvgen} contains
a sum of terms, each containing a product of 12 $\h_{ia}$. To obtain
the contribution of the diagram to a particular NMHV process we simply apply
the appropriate product of differential operators of total order 12.
This gives the value of the diagram in the original form \reef{mhvdi}.

In Sec.~4.2  we argued that the MHV vertex expansion of any particular
amplitude is unique and contains exactly the diagrams which come from the
recursion relation associated with any valid
 3-line shift  
$[m_1,m_2,m_3|$ which
satisfies the common index criterion.  A diagram is identified by specifying the
channel in which a pole occurs.  A 6-point amplitude $A_6(1,2,3,4,5,6)$ can
contain 2-particle poles in the channels $(12), (23), (34), (45), (56),$ or $(61)$,
and there can be 3-particle poles in the channels $(123), (234), (345)$.
However, different  6-point NMHV amplitudes contain different subsets of the
9 possible diagrams.  For example, the 6-gluon amplitudes with  helicity configurations $A_6(---+++)$ and $A_6(-+-+-+)$
each have one valid common index shift of the 3 negative helicity lines. In the first case,
there are 6 diagrams, since diagrams with poles  in the (45), (56) and (123) channels do not occur in the recursion relation, but all 9 possible diagrams
contribute to the recursion relation for the second case.  The amplitude
$\<A^1(1)A^{12}(2)A^{23}(3)A^{234}(4)A^{134}(5)A^4(6)\>$  for a process with 4 gluinos and 2 scalars is a more curious example;  its MHV vertex expansion
contains only one diagram with pole in the (45) channel.

We would like to define a universal generating function which contains the amplitudes for \emph{all} $n$-point NMHV amplitudes, such that any particular amplitude
is obtained by applying the appropriate 12th order differential operator.  It is natural to define the  generating function as
\be \lab{unigen}
\cf_n = \sum_I  \cf_{I,n}\,
\ee
in which we sum the generating functions \reef{nmhvgen} for \emph{all $n(n-3)/2$ possible}
diagrams that can appear in the MHV vertex expansion of 
$n$-point
amplitudes, for example all 9 diagrams listed above for 6-point
amplitudes.
If a particular diagram
$\bar{I}$   does not appear in the MHV vertex expansion of a given amplitude,
then the spin factor obtained by applying the appropriate
Grassmann differential operator to the generating function $\cf_{\bar{I},n}$
must vanish, leaving only the actual diagrams which contribute to the expansion.

To convince the reader that this is true, we first make an observation which follows from the way in which each $\cf_{I,n}$ is constructed starting from \reef{nmhvprod}.  We observe that the
result of the application of a Grassmann derivative $\cd^{(12)}$ of order 12 in the external $\h_{ia}$ to any $\cf_{I,n}$ is the same as the result of applying the operator
$\cd^{(16)} = \cd^{(12)} D_{I_L} D_{I_R}$ to the product in \reef{nmhvprod}.
If non-vanishing, this result is simply the product of the
spin factors for the left and right subdiagrams, so the contribution of the diagram $I$ to the amplitude corresponding to $\cd^{(12)}$ is correctly obtained.

We now show that  $\cd^{(12)}\cf_{\bar{I},n}$ vanishes when a diagram $\bar{I}$ does not contribute to the corresponding amplitude.
We first note that the amplitudes governed by $\cf_n$ are all NMHV. Thus they all have
overall $\h$-count 12, and $SU(4)$ invariance requires that each index value $a=1,2,3,4$ must
appear exactly 3 times among the external lines. Denote the lines which carry the index $a$ in the amplitude under study by $q_{1a},q_{2a},  q_{3a}$.
Consider a diagram $I$ and suppose that for every value of $a$ its two subamplitudes each contain at least one line from the set $q_{ka},~k=1,2,3.$
Then the diagram
$I$ appears in the MHV vertex expansion of the amplitude, and
the diagram contributes correctly to $\cd^{(12)}\cf_n$.
The other possibility is that there is a diagram $\bar{I}$
such that for some index value $b$
the 3 lines $q_{1b},q_{2b},  q_{3b}$  appear in only one subamplitude, say the left
subamplitude. Then the right subamplitude cannot be $SU(4)$ invariant. Its spin factor vanishes and the diagram does not contribute.

\setcounter{equation}{0}
\section{Spin state sums for loop amplitudes}
\lab{s:cuts}

Consider the $L$-loop amplitude shown in figure \ref{fig:gencut}. The evaluation of the $(L+1)$-line unitarity cut involves a sum over all intermediate states that run in the loops. The generating functions allow us to do such sums very efficiently, for any arrangements of external states as long as the left and right subamplitudes, denoted $I$ and $J$, are either MHV or NMHV tree amplitudes.

We begin by a general analysis of cut amplitudes of the type in figure \ref{fig:gencut}.
Assume that the full amplitude is N$^k$MHV. Then the total $\eta$-count is
$\sum_{\rom{ext}\, i} r_i = 4(k+2)$. Let the $\eta$-count of the $l$th loop state on the subamplitude $I$ be $w_l$; then that same line will have $\eta$-count $4-w_l$ on the subamplitude $J$. The total $\eta$-counts on the subamplitudes $I$ and $J$ are then, respectively,
\bea
  r_I = \sum_{\rom{ext}\,i \in I} r_i + \sum_{l=1}^{L+1} w_l\, ,~~~~~~
  r_J = \sum_{\rom{ext}\,j \in J} r_j + \sum_{l=1}^{L+1} (4-w_l) \, ,
\eea
so that
\bea
 \lab{rIrJ}
  r_I + r_J = \sum_{\rom{ext}\, i}  r_i + 4(L+1) = 4(k+L+3) \, .
\eea
Each subamplitude $I$ and $J$ must have an $\eta$-count $r_{I,J}$ which is a multiple of 4. If the overall amplitude is MHV and $L=1$, then \reef{rIrJ} gives $r_I+r_J = 16$, and the only possibility is that both subamplitudes $I$ and $J$ are MHV with $\eta$-counts 8 each. (Total $\eta$-count 4 is non-vanishing only for a 3-point anti-MHV amplitude; such spin sums are considered in section \ref{s:anti}.) Likewise, a 2-loop MHV amplitude has $r_I+r_J=20 = 8 + 12= 12+ 8$, so the intermediate state sum splits into MHV $\times$ NMHV plus NMHV $\times$ MHV.

The table in figure \ref{fig:gencut} summarizes the possibilities for MHV and NMHV loop amplitudes with $(L+1)$-line cuts. For each split, one must sum over all intermediate states; the tree generating functions allow us to derive new generating functions for cut amplitudes with all intermediate states summed.

\begin{figure}[t]
\begin{center}
 \includegraphics[width=7cm]{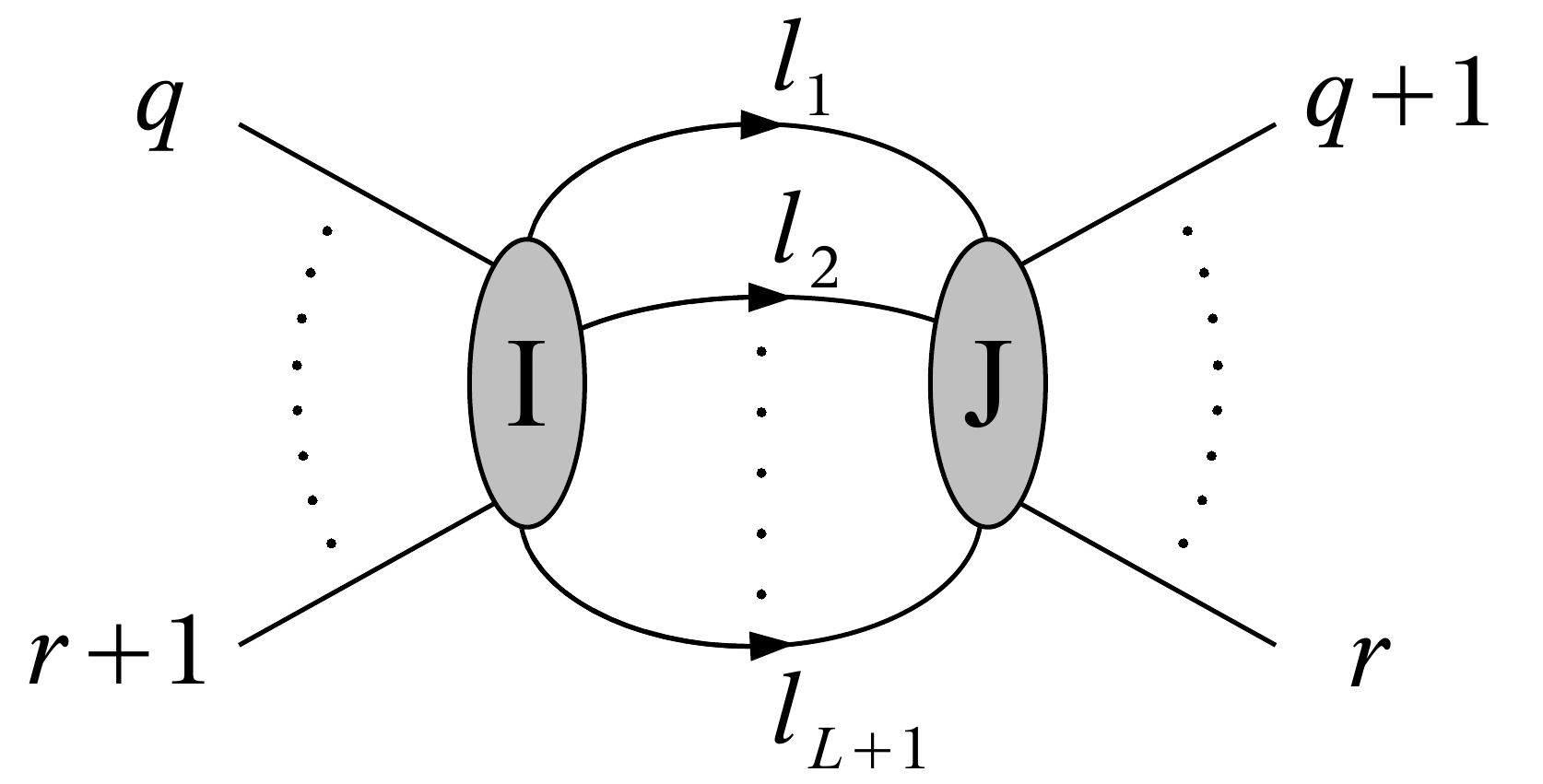}\\[5mm]
\begin{tabular}{|lccccc|}
  \hline \rule{0pt}{3ex}
  external && $L=1$  & $L=2$ & $L=3$ & $L=4$\\[1mm]
  \hline \rule{0pt}{7ex}
  MHV
  &$\to$& MHV $\times$ MHV
  & $\begin{array}{c} \rom{MHV} \times \rom{NMHV} \\  \rom{NMHV} \times \rom{MHV} \end{array}$
  & $\begin{array}{c} \rom{MHV} \times \rom{N^2MHV} \\  \rom{NMHV} \times \rom{NMHV} \\ \rom{N^2MHV} \times \rom{MHV}\end{array}$
  &$\begin{array}{c} \rom{MHV} \times \rom{N^3MHV} \\  \rom{NMHV} \times \rom{N^2MHV} \\ \rom{N^2MHV} \times \rom{NMHV} \\ \rom{N^3MHV} \times \rom{MHV} \end{array}$  \\[1mm]
  \hline \rule{0pt}{6ex}
  NMHV
  &$\to$& $\begin{array}{c} \rom{MHV} \times \rom{NMHV} \\  \rom{NMHV} \times \rom{MHV} \end{array}$
  & $\begin{array}{c} \rom{MHV} \times \rom{N^2MHV} \\  \rom{NMHV} \times \rom{NMHV} \\ \rom{N^2MHV} \times \rom{MHV}\end{array}$
  & etc& \\[1mm]
  \hline
\end{tabular}
\end{center}
\caption{N$^k$MHV loop amplitude evaluated by
 a
unitarity cut of $(L+1)$-lines. The sum over intermediate states involves all subamplitudes $I$ and $J$ with  $\eta$-counts $r_I$ and $r_J$ such that $r_I + r_J=4(k+L+3)$.
(For $L=1$ we assume that $I$ and $J$ each have more than one external leg, so that 3-point anti-MHV does not occur in the spin sum.)
}
\lab{fig:gencut}
\end{figure}

We outline the general strategy before presenting the detailed examples. Let $\cf_I$ and $\cf_J$ be generating functions for the subamplitudes $I$ and $J$ of the cut amplitude. To evaluate the cut, we must act on the product  $\cf_I\, \cf_J$ with the differential operators of all the external states $D^{(4k+8)}_\rom{ext}$ and of all the internal states
 $D_1 D_2 \cdots D_{L+1}$.
 The fourth order differential operators of the internal lines distribute themselves in all possible ways between $\cf_I$ and $\cf_J$ and thus automatically carry out the spin sum.
In \cite{BEF} it was shown how to evaluate the 1-loop MHV state sums when the external lines where all gluons. This was done by first acting with the derivative operators of the external lines, and then evaluating the derivatives for the loop states. We generalize the approach here to allow any set of external states of the $\cn = 4$ theory. This is done by postponing the evaluation of the external state derivatives, and instead carrying out the
the internal line Grassmann derivatives by converting them to Grassmann integrations. The result is a generating function
$\cf_\rom{cut}$ for the cut amplitude. It is defined as
\bea
 \cf_\rom{cut} & = &
 D_1 D_2 \cdots D_{L+1}
 ~\cf_I\, \cf_J \, .
\eea
The value of a particular cut amplitude is found by applying the external state differential operators $D^{(4k+8)}_\rom{ext}$ to $\cf_\rom{cut}$.

In the following we derive generating functions for unitarity cuts of 1-, 2- and 3-loop MHV and NMHV amplitudes.
Spin sums involving N$^2$MHV and N$^3$MHV subamplitudes for $L=3,4$ are carried out using anti-MHV and anti-NMHV generating functions in section \ref{s:anti}.


\subsection{1-loop intermediate state sums}

\subsubsection{1-loop MHV $\times$ MHV}
Consider the intermediate state sum in a 2-line cut 1-loop amplitude.
Let the external states be any $\cn=4$ states such that the full loop amplitude is MHV. By the analysis above, the subamplitudes $I$ and $J$ of the cut loop amplitude must then also be MHV.

We first calculate the intermediate spin sum and then include the appropriate prefactors. The state dependence of an MHV subamplitude is encoded in the
$\d^{(8)}$-factor of the MHV generating function. We will refer to the sum over spin factors as the ``spin sum factor'' of the cut amplitude. For the present case, the spin sum factor is
\bea
  D_1 D_2\,  \d^{(8)}(I) \, \d^{(8)}(J)
\eea
with $D_i = \prod_{a=1}^4\pa/\pa \eta_{i a}$ being the 4th order derivatives associated with the internal line $l_i$ and
\bea
  \lab{IJ}
  \begin{split}
  I &=&
  |l_1\> \eta_{1a} + |l_2\> \eta_{2a}
  + \sum_{\mathrm{ext}\; i \in I} |i\> \eta_{ia} \, , \\
  J &=&
  - |l_1\> \eta_{1a} - |l_2\> \eta_{2a}
  + \sum_{\mathrm{ext}\; j \in J} |j\> \eta_{ja} \, .
  \end{split}
\eea

We proceed by converting the $D_1 D_2$ Grassmann differentiations to integrations. Perform first the integration over $\eta_2$
to find \cite{ggk}
\bea
 \nonumber
 D_1 D_2\,  \d^{(8)}(I) \, \d^{(8)}(J)
 &=& \int d^4 \eta_1 \,  d^4 \eta_2 \;  \d^{(8)}(I) \, \d^{(8)}(J)\\
 &=& \d^{(8)}(I+J) \int d^4 \eta_1 \,
 \prod_{a=1}^4 \Big( \sum_{\mathrm{ext}\; j \in J} \<l_2\, j\> \, \eta_{ja}
   - \< l_2 \,l_1 \> \, \eta_{1a}\Big) \, .
\eea
The delta-function $\d^{(8)}(I+J)$ involves only the sum over external states and does therefore not depend on $\eta_1$.
The $\eta_1$-integrations picks up the $\eta_1$ term only, so we simply  get
\bea
 D_1 D_2\,  \d^{(8)}(I) \, \d^{(8)}(J)
 &=&
 \< l_1 \, l_2 \>^4\;
 \d^{(8)} \Big( \sum_{\mathrm{all\; ext}\; m} |m\>  \eta_{m a}\Big) \, .
\eea

If the external states are two negative helicity gluons $i$ and $j$ and the rest are positive helicity gluons, then, no matter where the gluons $i$ and $j$ are placed, we get $ \< l_1 \, l_2 \>^4  \< i \, j \>^4$, in agreement with (5.6) of \cite{bddk} and  (4.9) of \cite{BEF}.

Let us now include also the appropriate pre-factors in the generating function. For the MHV subamplitudes these are simply the cyclic products of momentum angle brackets. Collecting the cyclic product of \emph{external} momenta, we can write the full generating function for the MHV $\times$ MHV 1-loop generating function as
\bea
  \nonumber
  \cf^\text{1-loop}_{\mathrm{MHV} \times \mathrm{MHV}}
  &=&
  \frac{\< q, q+1 \> \<r, r+1 \> \<l_1 l_2 \>^2}
     {\< q\, l_1\> \< q +1, l_1\> \< r\, l_2\> \< r+1, l_2\>}\,
      \frac{1}{\prod_{\rom{ext}\; i} \<i,i+1\>} 
     ~\d^{(8)} \Big( \sum_{\mathrm{all\; ext}\; m} |m\>  \eta_{ma}\Big) \, ,
\eea
 or simply,
 \bea
 \boxed{~~~
  \cf^\text{1-loop}_{\mathrm{MHV} \times \mathrm{MHV}}
   ~=~
   \frac{\< q, q+1 \> \<r, r+1 \> \<l_1 l_2 \>^2}
     {\< q\, l_1\> \< q +1, l_1\> \< r\, l_2\> \< r+1, l_2\>}~
   \cf^\text{tree}_{\mathrm{MHV}}(\mathrm{ext}) \, . ~~~~}
\eea
Note that the state dependence of the cut MHV $\times$ MHV amplitude is included entirely in the MHV generating function, and all dependence on the loop momentum is in the prefactor.


\subsubsection{Triple cut of NMHV 1-loop amplitude: MHV $\times$ MHV $\times$ MHV}
\lab{s:triple}

\begin{figure}
\begin{center}
 \includegraphics[width=6cm]{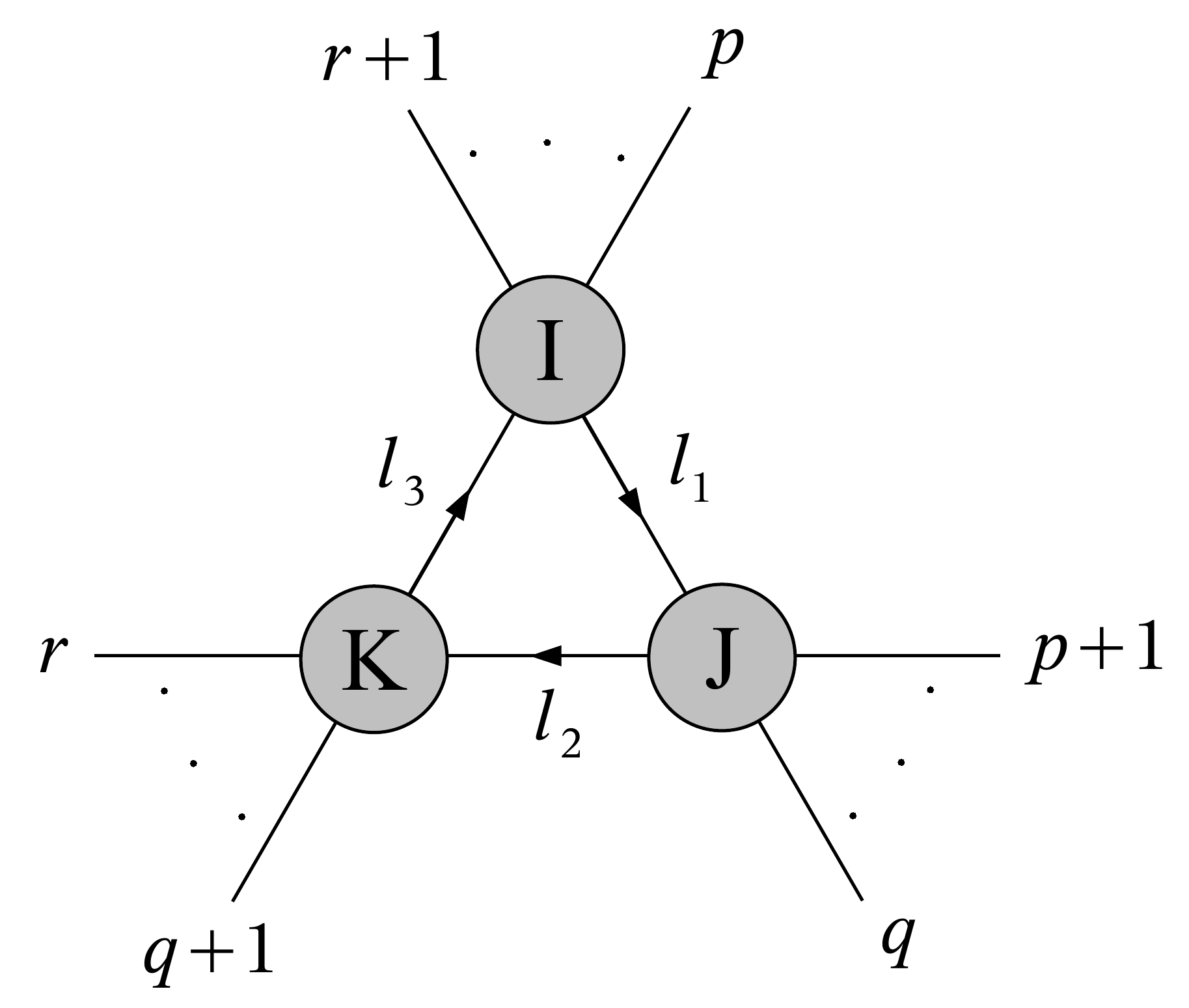}\\[5mm]
\end{center}
\vspace{-5mm}
\caption{Triple cut of NMHV 1-loop amplitude gives MHV subamplitudes $I$, $J$, and $K$.}
\lab{fig:triple}
\end{figure}

In this section we evaluate the intermediate state sum for a
1-loop NMHV amplitude with a triple cut, as illustrated in figure \ref{fig:triple}.
The triple cut is different from the cuts considered
 at the beginning of 
section \ref{s:cuts}. Its primary feature is that it gives three subamplitudes which are all MHV. To see this, note that the $\eta$-counts of the subamplitudes $I$, $J$ and $K$ are
\bea
  r_I = \sum_{\rom{ext}\; i \in I} r_i + w_1 + 4-w_3\, , ~~~
  r_J = \sum_{\rom{ext}\;j \in J} r_j + w_2 + 4-w_1\, , ~~~
  r_K = \sum_{\rom{ext}\;k \in K} r_k + w_3 + 4-w_2\, ,~
\eea
where the $r_i$ are the $\eta$-counts of the external states and $w_l$ and $4-w_l$ are the $\eta$-counts at each end of the internal lines. Since the full amplitude is NMHV, we have
\bea\label{rIrJrK}
  r_I + r_J + r_K= \sum_{\rom{all\; ext}\; i} r_i + 12 = 24 \, .
\eea
 We now assume that each subamplitude $I$, $J$, and $K$ has more than three legs and thus more than one external leg.
 Then~(\ref{rIrJrK})
has only one solution, namely $r_I=r_J=r_K=8$, so each subamplitude is MHV.

 Let us again first evaluate the spin sum and include the appropriate prefactors at the end.
The spin sum factor is calculated by letting the differential operators of the internal states act on the product of the three MHV generating functions for the subamplitudes.
 We have
\bea
 f^\mathrm{triple}
 &=& D_1 D_2 D_3 \big[\d^{(8)}(I)\; \d^{(8)}(J)\;  \d^{(8)}(K)\big]
\, ,
\eea
where
\bea
\nonumber
I &=& |l_1\>\h_{1a} - |l_3\>\h_{3a}
+\sum_{\rom{ext}\; i \in I} |i\>\h_{ia} \, ,\\
\lab{IJK}
J &=&  -|l_1\>\h_{1a} + |l_2\>\h_{2a}
+\sum_{\rom{ext}\; j \in J} |j\>\h_{ja}\, ,\\
\nonumber
K&=&  -|l_2\>\h_{2a} + |l_3\>\h_{3a}
+\sum_{\rom{ext}\; k \in K} |k\>\h_{ka}\, .
\eea
Again we convert the differentiations to integrations, and perform the integrations one at a time to find
\bea
\nonumber
  f^\mathrm{triple}
 &=&
  \int d^4 \eta_1 \,d^4 \eta_2 \, d^4 \eta_3 \;
  \d^{(8)}(I)\; \d^{(8)}(J)\; \d^{(8)}(K) \\
\nonumber
 &=&
 \int d^4 \eta_2 \, d^4 \eta_3 \;
   \d^{(8)}(I+J) \, \d^{(8)}(K)\,
   \prod_{a=1}^4 \Big( \sum_{\rom{ext}\; i \in I} \<l_1 i\> \eta_{ia}
   - \< l_1 l_3 \> \eta_{3a}\Big)\\ \nonumber
 &=&
 \d^{(8)}(I+J+K) \,
 \int d^4 \eta_3 \,
 \prod_{a=1}^4 \Big( \sum_{\rom{ext}\; i \in I} \<l_1 i\> \eta_{ia}
   - \< l_1 l_3 \> \eta_{3a}\Big)
  \Big( \sum_{\rom{ext}\; k \in K} \<l_2 k\> \eta_{ka}
   + \< l_2 l_3 \> \eta_{3a}\Big)\\
 &=&
\d^{(8)}\Big(\sum_\mathrm{all\; ext \; m} |m\> \eta_{mb}\Big)
\prod_{a=1}^4
\Big( - \sum_{\rom{ext}\; k \in K} \< l_3 l_1 \> \<l_2 k\> \eta_{ka}
   +   \sum_{\rom{ext}\; i \in I} \< l_2 l_3 \> \<l_1 i\> \eta_{ia}\Big) \, .
\eea
This is the generating function for the spin sum factor of the triple cut.\footnote{Note that using the overall $\d^{(8)}$ and the Schouten identity, the $\prod \sum$-factor can be rearranged cyclically as
\bea
  \nonumber
  - \sum_{\rom{ext}\; k \in K} \< l_3 l_1 \> \<l_2 k\> \eta_{ka}
   +   \sum_{\rom{ext}\; i \in I} \< l_2 l_3 \> \<l_1 i\> \eta_{ia}
 &=&
  - \sum_{\rom{ext}\; i \in I} \< l_1 l_2 \> \<l_3 i\> \eta_{ia}
   +   \sum_{\rom{ext}\; j \in J} \< l_3 l_1 \> \<l_2 j\> \eta_{ja} \\
 &=&
  - \sum_{\rom{ext}\; j \in J} \< l_2 l_3 \> \<l_1 j\> \eta_{ja}
   +   \sum_{\rom{ext}\; k \in K} \< l_1 l_2 \> \<l_3 k\> \eta_{ka} \, .
\eea}

If the external particles are all gluons with three negative helicity gluons $i',j',k'$ distributed on the cut with $i' \in I$, $j' \in J$, and $k' \in K$, then the triple cut spin sum factor is
\bea
  \nonumber
  D_{i'}D_{j'}D_{k'}\, f^\mathrm{triple}
 &=&
 D_{i'}D_{k'} \,
 \Big\{
   \Big[
   D_{j'} \d^{(8)}\Big(\sum_\mathrm{all\; ext \; m} |m\> \eta_{mb}\Big)
   \Big]
   \prod_{a=1}^4
\Big(  -\< l_3 l_1 \> \<l_2 k'\> \eta_{k'a}
   +  \< l_2 l_3 \> \<l_1 i'\> \eta_{i'a} + \dots \Big)
 \Big\} \\ \nonumber
 &=&
  D_{i'}D_{k'} \prod_{a=1}^4
\Big( \< j' i' \>  \eta_{i'a} +\< j' k' \>  \eta_{k'a} +\dots \Big)
 \Big(  -\< l_3 l_1 \> \<l_2 k'\> \eta_{k'a}
   +  \< l_2 l_3 \> \<l_1 i'\> \eta_{i'a} + \dots \Big) \\ \nonumber
  &=&
  \big(\< j' i' \> \< l_1 l_3 \> \<l_2 k'\>
    - \< j' k' \> \< l_2 l_3 \> \<l_1 i'\>\big)^4
  \\
  &=&
  \big(\< l_1 j' \> \<  l_3 i' \> \<k' l_2\>
    - \< l_2 j' \> \<  l_3 k' \> \< l_1 i'\>\big)^4 \, .
\eea
This agrees\footnote{The power in \cite{unexp,BEF} was $8$, not $4$, because the calculations were done in $\cn=8$ SG instead of $\cn = 4$ SYM.} with (4.23) of \cite{unexp} and (4.13) of \cite{BEF}.

To complete the calculation, we must include the appropriate prefactors. The full MHV $\times$ MHV $\times$ MHV triple cut 1-loop generating function is then
\bea
  \hspace{-2mm}
  \boxed{
  \begin{array}{rcl}
  \cf^\text{1-loop NMHV triple cut}_\rom{MHV^3} &=&
  {\displaystyle \frac{\< p, p+1 \> \< q, q+1 \> \<r, r+1 \>}
     {\< p\, l_1\> \< l_1, p+1\> \< q\, l_2\> \< l_2, q+1\>
     \< r\, l_3\> \< l_3, r+1\>}}
     \times  \;
   \cf^\text{tree}_{\mathrm{MHV}}(\mathrm{ext})
   \\[4mm]
   && \hspace{1mm}
   \times
    \prod_{a=1}^4
    \Big(   - \sum_{\rom{ext}\; k \in K} \< l_3 l_1 \> \<l_2 k\> \eta_{ka}
   +   \sum_{\rom{ext}\; i \in I} \< l_2 l_3 \> \<l_1 i\> \eta_{ia}  \Big) \, .
   \end{array}}
\eea
It is interesting to note that the structure of
 $\cf^\text{tree}_{\mathrm{MHV}}\times \prod_a \sum$
is very similar to the NMHV generating function for an MHV vertex diagram.


\subsection{MHV 2-loop state sum with NMHV $\times$ MHV}

As a first illustration of the application of the NMHV generating function, we calculate the intermediate state sum of a 3-line cut 2-loop MHV amplitude. The state sum splits into two separate cases NMHV $\times$ MHV and MHV $\times$ NMHV (see section \ref{s:cuts}). It suffices to derive an expression for the generating function of the NMHV $\times$ MHV state sum; from that the MHV $\times$ NMHV sum is easily obtained by relabeling momenta.

We express the NMHV subamplitude $I$ in terms of its MHV vertex expansion. We denote by $I_L$ each MHV vertex diagram in the expansion, and we also let $I_L$ and $I_R$ label the Left and Right MHV subamplitudes of the diagram. For each MHV vertex diagram $I_L \subset I$ we compute the spin sum factor
\begin{equation}
  f_{I_L} =
  D_1D_2D_3\Big[\Big(\delta^{(8)}(I)\prod_{a=1}^4 \sum_{i\in I_L}\<i P_{I_L} \>\eta_{ia}\Big)\delta^{(8)}(J)\Big]\,.
\end{equation}
The prefactors of the generating functions will be included later when we sum the contributions of all the diagrams.
We are free to define the \emph{left} MHV subamplitude $I_L$ to be the one containing \emph{either one or none} of the loop momenta. For definiteness, let us denote the loop momentum contained in $I_L$ by $l_\alpha$, the others by $l_\beta$, $l_\gamma$. (If $I_L$ does not contain any loop momentum, this assignment is arbitrary.) Since $l_\b,l_\g \notin I_L$ we get
\bea
  \nonumber
    f_{I_L}&=&
    \int d^4\eta_{\alpha} \, d^4\eta_{\beta} \, d^4\eta_{\gamma}
    \Bigl(\delta^{(8)}(I)\delta^{(8)}(J)
    \prod_{a=1}^4\sum_{i\in I_L}\<i P_{I_L} \>\eta_{ia}\Bigr)\\
    \nonumber
     &=& \delta^{(8)}\bigl(I+J\bigr) ~
    \int d^4\eta_{\alpha} \, d^4\eta_{\beta}
  \prod_{a=1}^4\Big( \sum_{i'\in I} \<i'\gamma\>\eta_{i'a}
  \Big)
  \Big(\sum_{i\in I_L}\<i P_{I_L} \>\eta_{ia}\Big)\\ \nonumber
    &=&  \delta^{(8)}\bigl(I+J\bigr)  ~
    \int d^4\eta_{\alpha} \, d^4\eta_{\beta}
  \prod_{a=1}^4\Big( \< \g \a \> \eta_{\a a} +
  \< \g \b \> \eta_{\b a} +\dots
  \Big)
  \Big(
   \sum_{{\rm ext}\,i\in I_L} 
  \<i P_{I_L} \>\eta_{ia} + \d_{l_\a \in I_L} \<\a P_{I_L}\> \eta_{\alpha a} \Big)\\
   &=&
    \delta_{l_\alpha\in I_L}
    \<\beta \gamma\>^4 \,\<\a P_{I_L}\>^4\,
 \d^{(8)} \Big( \sum_{\mathrm{all\; ext}\; k} |k\>  \eta_{ka}\Big)
    \, .   \lab{ssf1}
\eea
We have introduced a Kronecker delta $\delta_{l_\alpha\in I_L}$ which is 1 if $l_\alpha\in I_L$ and zero otherwise.
If $l_\alpha \notin I_L$, then none of the internal momenta connect to $I_L$. The calculation shows that such ``1-particle reducible'' diagrams do not contribute to the spin sum. This is a common feature of all spin sums we have done.

Including now the prefactors and summing over all MHV vertex diagrams $I_L \subset I$, the generating function for the cut 2-loop amplitude is
\bea
  \cf^\text{2-loop, $n$-pt}_{\rom{NMHV} \times \rom{MHV}}
  &=&
  \frac{1}{\prod_{j \in J} \<j,j+1\>}
  \sum_{I_L \subset I} \Khoze{I_L} \frac{1}{\prod_{i \in I} \<i,i+1\>}
  \, f_{I_L} \, ,
\eea
where $\Khoze{I_L}$ is the prefactor
\reef{Khoze}.
Separating out the dependence on the external states into an overall factor $\cf^\rom{tree}_\rom{MHV}(\rom{ext})$, we get
\bea
  \lab{F2loop}
 \cf^\text{2-loop, $n$-pt}_{\rom{NMHV} \times \rom{MHV}}
  &=&
  \cf^\rom{tree}_\rom{MHV}(\rom{ext}) ~
  \frac{\< q,q+1\> \<r,r+1\>}
  {\<q, l_1 \> \<l_1, q+1\> \< r, l_3 \> \< l_3 ,r+1\>
    \< l_1 \, l_2 \>^2\< l_2 \, l_3 \>^2}
    \sum_{I_L \subset I} \Khoze{I_L}\,(S.F.)_{I_L}~~~~~~~
\eea
with
\bea
  \lab{ssf2}
  (S.F.)_{I_L} &=& \<\beta \gamma\>^4 \, \<\alpha P_{I_L}\>^4
  ~ \d_{l_\a \in I_L ; ~l_\b,l_\g \notin I_L} \,.
\eea
Each term in the sum over $I_L \subset I$ depends on the reference spinor $|X]$ through the prescription $|P_{I_L}\> = P_{I_L} |X]$, but the sum of all diagrams must be $|X]$-independent.

\begin{figure}
\begin{center}
 \includegraphics[width=5.5cm]{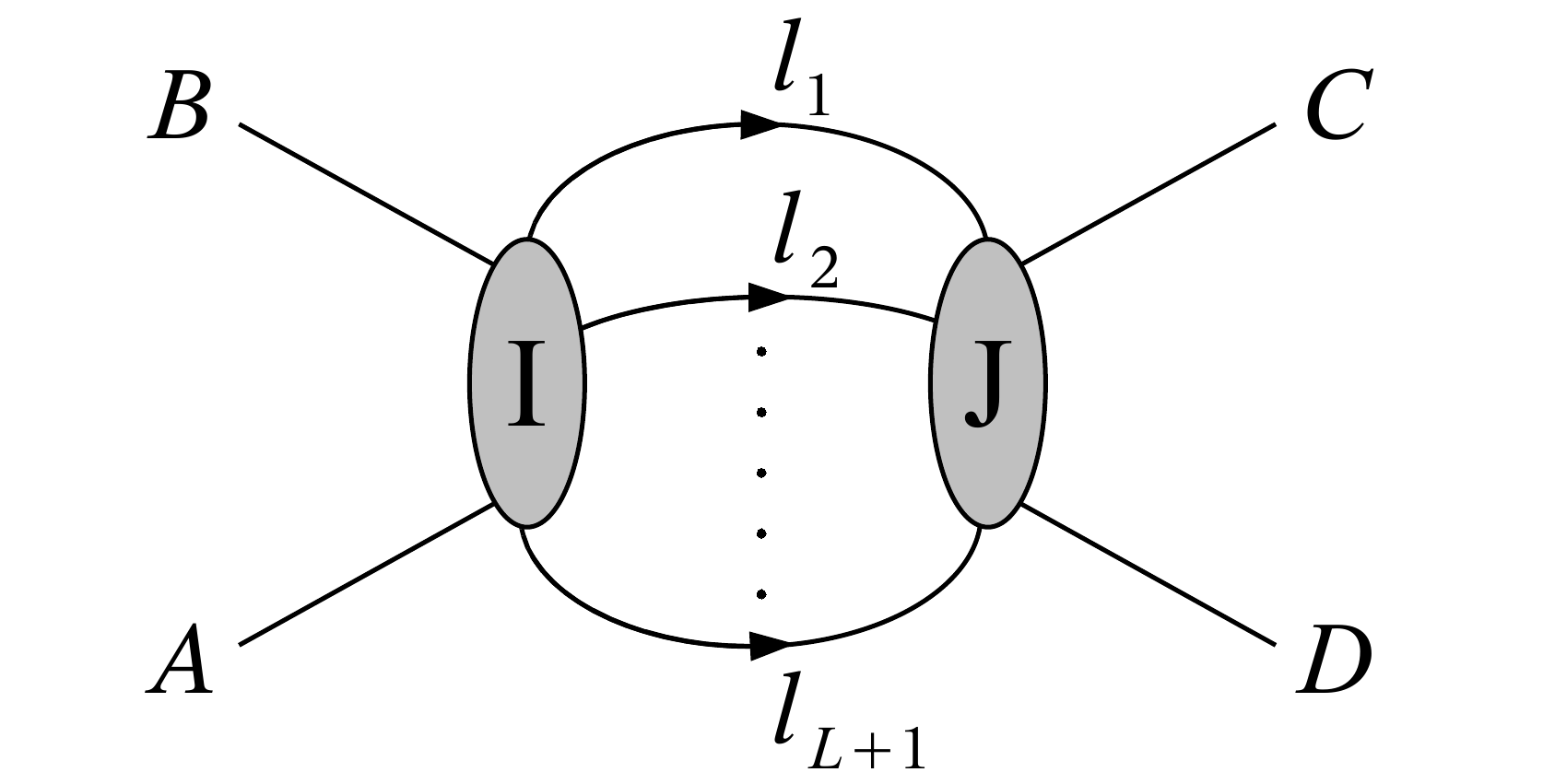}\\[5mm]
\end{center}
\vspace{-5mm}
\caption{4-point $L$-loop MHV amplitude with $(L+1)$-line cut.}
\lab{fig:ABCD}
\end{figure}

\subsubsection*{Example: 3-line cut of 4-point 2-loop amplitude}
 Let the external states be $A,B,C,D$, with $A,B$ on the subamplitude $I$ and $C,D$ on $J$, as shown in figure \ref{fig:ABCD} with $L=2$.
The
 subamplitude $I$
of the cut is a 5-point NMHV amplitude. Its MHV vertex expansion has five diagrams $(I_L | I_R)$, which we list with their spin sum factors:
\bea
  \nonumber
  &&(A, B | l_1,l_2,l_3)
  \hspace{3.32cm}\lra~~~~
  0\, ,\\
  &&
  (B,l_1|l_2,l_3,A) \, ,~~~
  (A,B,l_1|l_2,l_3)
  ~~~~~\lra~~~~ \<l_2 l_3\>^4 \< l_1 P_{I_L}\>^4\, ,\\ \nonumber
  &&
  (l_3,A|B,l_1,l_2) \, ,~~~
  (l_3,A,B| l_1,l_2)
   ~~~~~\lra~~~~ \<l_1 l_2\>^4 \< l_3 P_{I_L}\>^4 \, .
\eea
We have checked numerically that the sum $\sum_{I_L \subset I} \Khoze{I_L}\,(S.F.)_{I_L}$
is independent of the reference spinor $|X]$.

As a further check, let us assume that the two particles $C$ and $D$ are negative helicity gluons while the two particles $A$ and $B$ are positive helicity gluons. With the assumption that the cut is NMHV $\times$ MHV, there is only one choice for the internal particles: they have to be gluons, negative helicity coming out of the subamplitude $I$ and
 thus
positive helicity on $J$. So the spin sum only has one term, namely
\bea
  \nonumber
  &&
  A_5\big(A^+,B^+,l_1^-,l_2^-,l_3^-\big) \,
  A_5\big(C^-, D^-,-l_3^+,-l_2^+,-l_1^+\big) \\[2mm]
  &&~~~~=~
  \frac{[A\,B]^3}{[B\, l_1] [l_1\, l_2] [l_2 l_3] [l_3 \,A]}
  \frac{\<C\,D\>^3}
  {\< D \, l_3\> \<l_3 \, l_2 \> \< l_2 \, l_1 \>
   \< l_1\,C \> 
  } \, .
  \lab{a5a5}
\eea
This should be compared with the result of the spin sum \reef{F2loop} with the appropriate spin state dependence from $\cf^\rom{tree}_\rom{MHV}(\rom{ext})$. We have checked numerically that the results agree.

We can use this result to replace the spin sum over $I_L \subset I$ in \reef{F2loop} by the anti-MHV $\times$ MHV factor \reef{a5a5} and then write the full 4-point generating function in the simpler form
\bea
  \lab{F2loop4pt}
  \boxed{~~
 \cf^\text{2-loop, 4-pt}_{\rom{NMHV} \times \rom{MHV}}
 ~=~
 \frac{[A\,B]^3 \<D\,A\> \< A\, B \> \<B\,C\>}{\<C|l_1|B] \<D|l_3|A] P_{l_1 l_2}^2 P_{l_2 l_3}^2} ~ \cf^\rom{tree}_\rom{MHV}(\rom{ext})
   \, .~}
\eea
We have checked numerically the agreement between \reef{F2loop} and
 \reef{F2loop4pt} for 4 external lines.
  The 
 generating function \reef{F2loop4pt} gives the correct result for any MHV choice of 4 external states.


\subsection{MHV 3-loop state sum with NMHV $\times$ NMHV}

Consider the NMHV $\times$ NMHV part of the 3-loop spin sum.
We express the $I$ and $J$ subamplitudes in terms of their MHV vertex expansions; thus in the intermediate state sum we must sum over all products of MHV vertex diagrams $I_L \subset I$ and $J_L \subset J$. We first compute the spin sum factor associated with such a product, then include the necessary prefactors in order to get a general expression for the intermediate state sum.

For each MHV vertex diagram of the subamplitudes $I$ and $J$, there is a freedom in choosing which MHV vertex we call ``left''. This always allows us to choose $I_L$ and $J_L$ such that neither contains the internal momentum line $l_4$. This is a convenient choice for performing the $\eta_4$ integration first and then evaluating the three other $\eta$-integrations. The spin sum factor for a product of MHV vertex diagrams $I_L$ and $J_L$ with $l_4 \notin I_L \cup J_L$ is then
\bea
  \nonumber
  &&\hspace{-2cm}
  D_1 D_2 D_3 D_4~
  \cf^\text{tree-diagram}_\text{NMHV}(I_L)
  ~ \cf^\text{tree-diagram}_\text{NMHV}(J_L)
   \\[2mm]  \nonumber
  &=&D_1 D_2 D_3 \int d^4\eta_4~
  \Big(\delta^{(8)}(I)\prod_{a=1}^4 \sum_{i\in I_L}\<i P_{I_L}\> \eta_{ia}\Big)~
  \Big(\delta^{(8)}(J)\prod_{b=1}^4 \sum_{j\in J_L}\<j P_{J_L}\> \eta_{jb}\Big)\\
  \nonumber
  &=&
  \delta^{(8)}(I+J)~
  D_1 D_2 D_3\,
  \prod_{a=1}^4
  \Big( \< l_1 l_4 \> \eta_{1a}
    +  \< l_2 l_4 \> \eta_{2a}
    +  \< l_3 l_4 \> \eta_{3a} + \dots \Big) \\
  \nonumber
   &&\hspace{2cm}
    \Big( \d_{l_1 \in I_L} \< l_1 P_{I_L} \> \eta_{1a}
    +  \d_{l_2 \in I_L} \< l_2  P_{I_L}  \> \eta_{2a}
    +  \d_{l_3 \in I_L}  \< l_3  P_{I_L} \> \eta_{3a} + \dots \Big) \\
  \nonumber
   &&\hspace{2cm}
    \Big( \d_{l_1 \in J_L} \< l_1 P_{J_L} \> \eta_{1a}
    +  \d_{l_2 \in J_L} \< l_2  P_{J_L}  \> \eta_{2a}
    +  \d_{l_3 \in J_L}  \< l_3  P_{J_L} \> \eta_{3a} + \dots \Big) \\[2mm]
   &=&
  \delta^{(8)}(I+J)~\text{(s.s.f.)}_{I_L,J_L} \, ,
\eea
where
\bea
  \lab{ssf3}
  \text{(s.s.f.)}_{I_L,J_L}
 &=&
 \left[
 \det
 \left(
   \begin{array}{ccc}
     \< l_1 l_4 \>
     &  \< l_2 l_4 \>
     &  \< l_3 l_4 \> \\
     \d_{l_1 \in I_L} \< l_1 P_{I_L} \>
     &  \d_{l_2 \in I_L} \< l_2  P_{I_L}  \>
     &  \d_{l_3 \in I_L}  \< l_3  P_{I_L} \> \\
     \d_{l_1 \in J_L} \< l_1 P_{J_L} \>
    &  \d_{l_2 \in J_L} \< l_2  P_{J_L}  \>
    &  \d_{l_3 \in J_L}  \< l_3  P_{J_L} \>
   \end{array}
 \right)
 \right]^4.
\eea

We must sum over all diagrams including the appropriate prefactors. There are $W_{I_L}$ and  $W_{J_L}$ factors \reef{details} from the two MHV vertex expansions, as well as cyclic products. With momentum labels $q$, $q+1$ etc as in figure \ref{fig:gencut} we can write the NMHV $\times$ NMHV part of the full 4-line cut 3-loop MHV amplitude as
\bea
  \boxed{
  \begin{array}{rcl}
  \cf_\text{NMHV $\times$ NMHV}^\text{3-loop,$n$-pt}
  &=& \displaystyle
  -\frac{\< q,q+1\> \<r,r+1\>}
  {\<q, l_1 \> \<l_1, q+1\> \< r, l_4 \> \< l_4 ,r+1\>
    \< l_1 \, l_2 \>^2\< l_2 \, l_3 \>^2\< l_3 \, l_4 \>^2}
  \\[5mm]
  &&\hspace{3cm} \displaystyle
  \times~ \cf^\rom{tree}_\rom{MHV}(\rom{ext}) ~
    \sum_{I_L \subset I,\; J_L \subset J} \Khoze{I_L}\Khoze{J_L}
    \text{(s.s.f.)}_{I_L,J_L}\, .
    \end{array}}
\eea
The product of any MHV vertex diagrams $I_L$ and $J_L$ involve two independent reference spinors $X_I$ and $X_J$ from the internal momentum prescriptions
$|P_{I_L}\> = P_{I_L} |X_I]$ and $|P_{J_L}\> = P_{J_L} |X_J]$, but the sum over all diagrams must be independent of both reference spinors.

Consider the 4-point 3-loop amplitude.
Let  the external states be $A,B,C,D$, with $A,B$ on the subamplitude $I$ and $C,D$ on $J$, as in figure \ref{fig:ABCD}. The NMHV subamplitudes $I$ and $J$ are 6-point functions, so their MHV vertex expansions involve a sum of 9 diagrams. For the subamplitude $I$ these diagrams are listed as $(I_L | I_R)$:
\bea
  \nonumber
  &&
  (AB \, | \, l_1 l_2 l_3 l_4)\, ,~~~~
  (B l_1  \, | \,  l_2 l_3 l_4 A)\, ,~~~~
  (l_1 l_2  \, | \,  l_3 l_4 A B)\, , \\
  &&
  (l_2 l_3  \, | \,  l_4 A B l_1)\, ,~~~~
  (A B l_1 l_2  \, | \,  l_3  l_4)\, ,~~~~
  (B l_1 l_2 l_3  \, | \,  l_4 A)\, , \\ \nonumber
  &&
  (AB l_1  \, | \,  l_2 l_3 l_4)\, ,~~~~
  (B l_1  l_2  \, | \,  l_3 l_4 A)\, ,~~~~
  (l_1 l_2  l_3  \, | \,  l_4 A B)\, .
\eea
For the subamplitude $J$, replace $A, B$ by $D, C$ to find $(J_L|J_R)$.
 (This gives reverse cyclic order on $J$.)
 Recall that we are assuming $l_4 \notin I_L,\,J_L$.

In some cases, the spin sum factor for product of diagrams $(I_L|I_R) \times (J_L|J_R)$ can directly be seen to vanish. For instance, if no loop momenta are contained in $I_L$ or $J_L$, then a row in the matrix \reef{ssf3} vanishes, and hence $(\rom{s.s.f.})_{I_L,J_L} = 0$. It follows that the diagrams $(AB \, | \, l_1 l_2 l_3 l_4)$ and $(DC \, | \, l_1 l_2 l_3 l_4)$ do not contribute to the spin sum.
Another non-contributing case is when $I_L$ and $J_L$ each contain only one loop momentum $l_i$ which is common to both. Then a $2 \times 2$ submatrix of  \reef{ssf3} vanishes, and hence $(\rom{s.s.f.})_{I_L,J_L} = 0$. This means that products such as $(B l_1  \, | \,  l_2 l_3 l_4 A) \times (DC l_1  \, | \,  l_2 l_3 l_4)$ vanish.
Finally, it may be noted that if $l_1,l_2,l_3 \in I_L \cap J_L$, then the determinant \reef{ssf3} vanishes thanks to the Schouten identity.
These observations are general and apply for any number of external legs to reduce the number of terms contributing in the sum over all products of MHV vertex diagrams.  For the case of 4 external momenta, the number of contributing diagrams are thus reduced from $9^2=81$ to $8^2-4 -4=56$.

We have verified numerically for the 4-point amplitude that the sum of all diagrams is independent of both reference spinors.


\setcounter{equation}{0}
\section{Anti-MHV and anti-NMHV generating functions and spin sums} \lab{s:anti}

In the previous section we have evaluated spin sums for unitarity cuts which  involved MHV and NMHV subamplitudes.\footnote{ By conjugation, these results apply quite directly to  cuts only involving anti-MHV and anti-NMHV subamplitudes.} However, the table in Fig.~\ref{fig:gencut} shows that
this is not enough.   The unitarity cut
at  loop order $L=2,3$ includes the product of MHV and N$^2$MHV
amplitudes,
and N$^3$MHV is needed at 4-loop order.
Our method would then require the generating functions for
N$^2$MHV and N$^3$MHV amplitudes (see \cite{efk2} for their construction). However, the situation is
also workable if these
amplitudes have a small number of external lines. For example,
if we are interested in the 4-line cut of a $3$-loop
 $4$-point function, then the N$^2$MHV amplitudes we need are
6-point functions, and these are the complex conjugates of MHV amplitudes, usually called anti-MHV amplitudes. To evaluate their contribution   to the intermediate state sum we need an anti-MHV generating function expressed in terms of the original $\h_{ia}
$ variables, so we can apply our integration techniques. Thus we
first describe a general method to construct
anti-N$^k$MHV generating functions from N$^k$MHV generating functions and use it to find explicit expressions for the anti-MHV and anti-NMHV cases (section~\ref{s:pre}).
Then we apply these generating functions to evaluate several examples
of unitarity sums in which  (N)MHV amplitudes occur on one side of the cut and anti-(N)MHV amplitudes on the other (section ~\ref{s:antiapp}).
The most sophisticated example is the intermediate state sum for a $5$-line cut of a $4$-loop $4$-point function.

\subsection{Anti-generating functions}
\lab{s:pre}

An N$^k$MHV amplitude has external states whose $\eta$-counts $r_i$ add up to a total of $4(k+2)$. The total $\eta$-count is matched in the generating function, which must be a sum of monomials of degree $4(k+2)$ in the variables $\h_{ia}$.
The states of the conjugate anti-N$^k$MHV amplitude have $\eta$-counts $4-r_i$, so the total $\eta$-count for $n$-point amplitudes is $4(n-(k+2))$. Thus  anti-N$^k$MHV generating functions must contain monomials of degree $4(n-(k+2))$.
For example, the 3-point anti-MHV generating function has degree 4, and the 6-point anti-MHV and 7-point anti-NMHV cases both have degree 16.

The $\h$-count requirement is nicely realized if we define the anti-N$^k$MHV generating function as  \cite{sok2} the Grassmann Fourier transform  of the conjugate of the corresponding N$^k$MHV generating function.  Given a set
of
 $N$
Grassmann variables $\theta_I$ and their formal adjoints $\bar\theta^I$,
the Fourier transform of any function $f(\bar\theta^I)$  is defined as
\be \lab{gft}
\hat{f}(\theta_I)  \equiv \int d^N\bar{\theta} \exp(\theta_I \bar\theta^I)\,f(\bar\theta^I) \,.
\ee
Any $f(\bar\theta^I)$  is a sum of monomials of degree
$M \le N$, e.g.
$\bar\theta^J\cdots\bar\theta^K$, which can be ``pulled out" of the integral and
expressed as derivatives,  viz.
\bea \lab{plout}
 \int d^N\bar{\theta}\, \exp(\theta_I \bar\theta^I)\, \bar\theta^J\cdots\bar\theta^K&=&
 (-)^N\,
 \frac{\pa}{\pa \theta_J }\cdots\frac{\pa}{ \pa\theta_K}  \int d^N\bar{\theta}\, \exp(\theta_I \bar\theta^I)\nonumber\\
&=& \frac{\pa}{\pa \theta_J }\cdots\frac{\pa}{ \pa\theta_K}\prod_{I=1}^N \theta_I\nonumber\\
&=& \frac{1}{(N-M)!}
 \e^{J\dots K I_{M+1} \dots I_N}\th_{I_{M+1}} 
\ldots\th_{I_N}\,.
\eea

The procedure to convert an N$^k$MHV $n$-point generating function into an anti-N$^k$MHV generating function uses conjugation followed by the Grassmann Fourier transform.  The conjugate of any function\footnote{For simplicity it is assumed that the only complex numbers
contained in $f$ are the spinor components of $|i\>$ and $|i]$.}
$f(\<ij\>,[kl],\h_{ia})$ is defined as $f([ji],\<lk\>, \bar{\h}_i^a)$, including  reverse order of Grassmann monomials.
Evaluation of the Fourier transform
\be\lab{gft2}
\hat{f} \equiv \int \prod_{i,a} d\bar\h_i^a \,\exp\Bigl(\,\sum_{b,j}\h_{jb}\bar\h_j^b\Bigr)\,
f([ji],\<lk\>, \bar{\h}_i^a)\,,
\ee
is then equivalent to the following general prescription:
\begin{enumerate}
\item Interchange all angle and square brackets: $\< ij \> \lra [ji]$.\vspace{-0.15cm}
\item Replace $\eta_{ia} \; \to \;\pa_i^a \;= \;\frac{\pa}{\pa \eta_{ia}}$.\vspace{-0.25cm}
\item Multiply the resulting expression by $\prod_{a=1}^4 \prod_{i=1}^n \eta_{ia}$ from the right.
\end{enumerate}

We first apply this to find an anti-MHV generating function. We will
 confirm that
the result is correct by showing that it solves the SUSY Ward identities. We will then apply the prescription to find an anti-NMHV generating function.

\subsubsection{Anti-MHV generating function}
\lab{s:Amhv}

Applied  to the conjugate of the MHV generating function \reef{n4gen} (with \reef{del8}), the prescription  gives the anti-MHV generating function\footnote{We omit an overall factor $(-1)^n$ in $\bar{F}_n$. This has no consequence for our applications in spin sums.}
\bea\label{barFn0}
 \bar{F}_n
  &=&
  \frac{1}{\prod_{i=1}^n [i,i+1]} \frac{1}{2^4}
  \prod_{a=1}^4 \sum_{i,j = 1}^n \;[i j]\; \pa^a_i \pa^a_j ~
  \eta_{1a} \cdots \eta_{na} \, .
\eea
Evaluating the derivatives as in \reef{plout} we can write this as
\bea
  \lab{barFn}
  \bar{F}_n
  &=&
  \frac{1}{\prod_{i=1}^n [i,i+1]} \frac{1}{(2\, (n-2)!)^4}
   \prod_{a=1}^4
   \sum_{k_1,\ldots,k_n} 
   \eps^{k_1 k_2 \cdots k_n}\, [k_1 k_2]\, \eta_{k_3 a} \cdots \eta_{k_n a}
\eea
The sum is over all external momenta $k_i \in \{ 1,2,\dots,n\}$.

To confirm
that \reef{barFn} is correct we show that $\bar{F}_n$ obeys the SUSY Ward identities and produces the correct  all-gluon anti-MHV amplitude $A_n(1^+ 2^+ 3^-  \dots  n^-)$. This is sufficient because the SUSY Ward identities  have a unique solution \cite{BEF} in the MHV or anti-MHV sectors. Any (anti-)MHV amplitudes can be uniquely written as a spin factor times an $n$-gluon amplitude.
The desired $n$-gluon amplitude is obtained by applying the product
$D_3\dots D_n$ of 4th order operators of \reef{opcorr} for the $(n-2)$
 negative
gluons to the generating function \reef{barFn}.  It is easy to obtain the expected result
\bea
  \nonumber
  D_3 \cdots D_n \bar{F}_n
  &=&
   \frac{1}{\prod_{i=1}^n [i,i+1]} \frac{1}{(2\, (n-2)!)^4}
   \prod_{a=1}^4
   \sum_{k_1,k_2}
   (n-2)! \, \eps^{k_1 k_2 3 4 \dots n}\, [k_1 k_2] \\
   &=&
   \frac{[12]^4}{\prod_{i=1}^n [i,i+1]} \, .
\eea

The supercharges which act on generating functions are \cite{BEF}
\be \lab{such}
Q^a \,=\, \sum_{i=1}^n [i | \, \pa_i^a\,,~~~~~
\tQ_a\,=\,\sum_{i=1}^n |i\>\,\h_{ia}\,.
\ee
Ward identities are satisfied if
 $Q^a$ and $\tQ_a$ annihilate $\bar{F}_n$. Formally this requirement is satisfied by the Grassmann Fourier transform, but we find the following
 direct proof instructive.  We compute
\bea
  Q^a \bar{F}_n
  &\propto&
   \sum_{k_1,k_2,i,k_4,\ldots,k_n}  
  [k_1 k_2] [i| \; \eps^{k_1 k_2 i k_4 \cdots k_n} \,
  \eta_{k_4 a} \cdots \eta_{k_n a}
  ~=~ 0
\eea
by the Schouten identity. The argument for $\tQ_a$ is slightly more involved. First write
\bea \lab{sch1}
  \tQ_a \bar{F}_n
  &\propto&
  \sum_{i,k_1,\dots,k_n} |i\> [k_1 k_2]
  \eps^{k_1 k_2 k_3 k_4 \cdots k_n} \,
  \eta_{i a}\, \eta_{k_3 a}  \cdots \eta_{k_n a} \, .
\eea
Note that the product of $\eta$'s is nonvanishing only when $i \notin \{ k_3,\dots,k_n\}$, i.e.~when $i$ is $k_1$ or $k_2$. Thus
\bea \lab{cons3}
  \nonumber
  \tQ_a \bar{F}_n
  &\propto&
  -2 \sum_{k_1,\dots,k_n} |k_2\> [k_2 k_1]
  \eps^{k_1 k_2 k_3 k_4 \cdots k_n} \,
  \eta_{k_2 a}\, \eta_{k_3 a}  \cdots \eta_{k_n a} \\
  &=&
  -2 (n-2)! \sum_{k_1=1}^n\bigg(
   \prod_{i \ne k_1} \eta_{i a} 
  \bigg)
  \Big( \sum_{k_2=1}^n  |k_2\> [k_2 k_1] \Big)
  ~=~ 0 \, ,
\eea
due to momentum conservation. For given $k_1$ the product of $(n-1)$ factors of $\eta_{l_i a}$'s with $l_i \ne k_1$ is the same for all choices of $k_2$ and it was therefore taken out of the sum over $k_2.$
This completes the proof that \reef{barFn} produces all $n$-point anti-MHV amplitudes correctly.

For $n=3$, the generating function \reef{barFn} reduces to the anti-MHV 3-point amplitudes
\bea
   \bar{F}_3
   &=&
   \frac{1}{[12][23][31]}
   \,\prod_{a=1}^4 
   \Bigl( [12]\eta_{3a} + [31] \eta_{2a} + [23] \eta_{1a} \Bigr) \, , 
\eea
recently presented in \cite{qmc}.

An alternative form of the anti-MHV generating function can be given for $n \ge 4$. It is more convenient for calculations because it contains the usual  $\d^{(8)} (\sum_{i=1}^n |i\> \eta_{ia})$ as a factor.
The second factor requires the selection of two special lines, here chosen to be 1 and 2. The alternate form reads
\bea
  \lab{AbarFn}
  \bar{F}_n
  \,=\,
  \frac{1}{\<12\>^4\, \prod_{i=1}^n [i,i+1]} \frac{1}{(2(n-4)!)^4}
  \; \d^{(8)} (\sum_{i=1}^n |i\> \eta_{ia}) ~
   \prod_{a=1}^4 \sum_{k_3,\dots,k_n}
   \eps^{1 2 k_3 \cdots k_n}\,
   [k_3 k_4]\, \eta_{k_5 a} \cdots \eta_{k_n a} \, .~~
\eea
Arguments very similar to the ones above show that \reef{AbarFn}
 satisfies the Ward identities
and produces the correct gluon amplitude
$A_n(1^- 2^-3^+4^+5^-\dots n^-)$. Since these requirements have a unique realization, the two forms \reef{barFn} and \reef{AbarFn} must coincide.

For $n=4,5$ the anti-MHV generating function \reef{AbarFn}  reduces to the ``superamplitudes'' recently presented in \cite{sok2}. It is worth noting that for $n=4$, any MHV amplitude is also anti-MHV; using momentum conservation it can explicitly be seen that the anti-MHV generating function \reef{barFn}, or in the form \reef{AbarFn},  is equal to the MHV generating function for $n=4$.

\subsubsection{Anti-NMHV generating function}
\lab{s:Anmhv}

Any anti-NMHV
 $n$-point amplitude $I$
of the $\cn =4$ theory has an anti-MHV vertex
 expansion,
which is justified by the validity of the MHV vertex expansion of the conjugate NMHV amplitude. For each diagram of the
expansion we use the conjugate of the CSW prescription, namely
\bea
  |P_{I_L}] = P_{I_L} |X\> \, .
\eea
This involves a reference spinor $|X\>$. The sum of all diagrams is independent of $|X\>$.

We will obtain the anti-NMHV generating function by applying the prescription above to the conjugate of the NMHV generator \reef{nmhvgen}.  This prescription  directly gives
\bea\lab{barNFn0}
  \bar{\cf}_{n,I_L}
  &=&
  \frac{1}{\prod_{i=1}^n [i,i+1]} \; \overline{W}_{I_L}\; 
  \frac{1}{2^4}
  \prod_{a=1}^4 \sum_{i,j\in I}  \sum_{k \in I_L}
  [i j] [ P_{I_L}\,k] \; \pa_{i}^a\pa_{j}^a\pa_{k}^a\; \eta_{1a} \cdots \eta_{na} \, ,
\eea
where $\overline{W}_{I_L}$ 
is obtained from $W_{I_L}$ in \reef{details} by exchanging angle and square brackets.

Carrying out the differentiations and relabeling summation indices
gives the desired result:
\be\lab{barNFn}
  \bar{\cf}_{n,I_L}
   ~=~ 
  \frac{1}{\prod_{i=1}^n [i,i+1]} \; \overline{W}_{I_L} 
  \frac{1}{(2\, (n-3)!)^4}
  \prod_{a=1}^4
  \sum_{k_1 \in I_L}\;
  \sum_{k_2,\dots,k_n \in I}
  [P_{I_L} k_1] [k_2 k_3] \; \eps^{k_1 k_2 \dots k_n}\;
  \eta_{k_4 a} \cdots \eta_{k_n a} \, . 
\ee
The factor $1/((n-3)!)^4$ compensates
 for 
the overcounting produced by the contraction of the Levi-Civita symbol with the products of $\eta$'s.
The expression \reef{barNFn} contains a hidden factor of $\d^{(8)}(\mathrm{ext})$ and can be written
\begin{equation}
\bar{\cf}_{n,I_L} ~=~
    \frac{\overline{W}_{I_L} \; \d^{(8)}\big(\sum_{i=1}^n |i\> \eta_{ia}\big)} 
    { (2 (n-5)!)^{4}\<12\>^{4}\prod_{i=1}^n [i,i+1]}
  \prod_{a=1}^4
  \sum_{k_3 \in I_L}
  \sum_{k_4,\dots,k_n \in I}\!\!\!\!\! \eps^{12k_3 k_4\dots k_n}
   [k_3\, P_{I_L}]  [k_4 k_5]\, \eta_{k_6 a} \cdots \eta_{k_n a}\,.
    \lab{antiNMHV}
\end{equation}
It is not trivial to show that \reef{antiNMHV} follows from \reef{barNFn}. We present the proof in \cite{efk2}.

In analogy with Sec.~4.3 the universal anti-NMHV generating function is the sum
\be \lab{fullgen}
\bar{\cf}_n \,=\,
 \sum_{I_L \subset I}
\bar{\cf}_{n,I_L}
\ee
over all possible diagrams.

One check that this result is correct is to show that the SUSY charges
of \reef{such} annihilate $\bar{\cf}_n$.
This check can be carried out, but it is not a complete test that \reef{fullgen} is correct because the SUSY Ward identities do not have a unique solution in the
NMHV or anti-NMHV sectors.  See \cite{gris} or \cite{BEF}.
For this reason we show
in Appendix \ref{appantiMHVVE} that \reef{barNFn0} is obtained for any diagram starting from the product of anti-MHV generating functions for the left and right  subamplitudes.   Essentially we obtain \reef{barNFn0} by
 the
complex conjugate of the
process which led from \reef{etaint} to \reef{nmhvgen}. It then follows
that the application of external line derivatives (of total order $4(n-3)$)
to \reef{fullgen} produces the correct ``anti-MHV vertex expansion" of
the corresponding anti-NMHV amplitude.

\subsection{Anti-generating functions in intermediate spin sums}
\label{s:antiapp}

With the anti-MHV and anti-NMHV generating functions we can complete the unitarity sums for 3- and 4-loop 4-point amplitudes.

\subsubsection{$L$-loop anti-MHV $\times$ MHV spin sum}

Consider the $(L+1)$-line unitarity cut of an $L$-loop MHV amplitude, as in figure \ref{fig:gencut}. The intermediate spin sum will include a sector where one subamplitude is N$^{L-1}$MHV and the other is
MHV. We assume that the full amplitude has a total of 4 external legs, with 2  on each side of the cut, as in figure \ref{fig:ABCD}, so the tree subamplitudes have $L+3$ legs. Then\footnote{An $(L+3)$-point anti-MHV amplitude has $\eta$-count $4(n-(k+2))=4(L+1)$, so it is N$^{L-1}$MHV.} the N$^{L-1}$MHV subamplitude is anti-MHV and we can apply our anti-MHV generating function to obtain the spin sum.

The spin sum is
\bea
  \nonumber
  F_\text{anti-MHV $\times$ MHV}^{L\text{-loop MHV 4-point}} &=&
  D_1 \cdots D_{L+1} ~\bar{F}_{L+3}(I) ~F_{L+3}(J) \\[1mm]
  &=&
  D_1 \cdots D_{L+1} ~
  \frac{1}{\prod_{i\in I} [i,i+1]} ~\widetilde{\d^{(8)}(I)}
 ~
 \frac{1}{\prod_{j\in J} \<j,j+1\>} ~\d^{(8)}(J)
  \, ,
\eea
where
\bea
  \widetilde{\d^{(8)}(I)}
  \equiv
  \frac{1}{\<l_1 l_2\>^4} \;\frac{1}{(2(L-1)!)^4}
 \,\delta^{(8)}(I)\,
    \Big( \prod_{a=1}^4
      \sum_{k_3,\dots,k_{L+3}} [k_3 k_4] \;
      \eps^{l_1 l_2 k_3 \cdots k_{L+3}}
      \;\eta_{k_5a} \cdots \eta_{k_{L+3} a}
    \Big)\, 
\eea
is obtained from the Fourier transform; we
use the form \reef{AbarFn} for the anti-MHV generating function, selecting the loop momenta $l_1$ and $l_2$ as the two special lines. Then, focusing on the spin sum factor only, we have
\bea
  \text{s.s.f.} 
  &=& D_1 \cdots D_{L+1}\;
  \widetilde{\d^{(8)}(I)} \; \d^{(8)}(J) \, .
 \eea
Converting the $\eta_1$ differentiation to integration we find
\bea
  \nonumber
  \text{s.s.f.} 
  &=&
  \d^{(8)}(I+J)~
  D_2 \cdots D_{L+1}
  \Big\{
    \frac{1}{\<l_1 l_2\>^4} \frac{1}{(2(L-1)!)^4}\;
    \prod_{a=1}^4
      \big( \dots + \<l_1 l_2 \> \eta_{2 a}+\dots\big)     \\
      \nonumber
   &&\hspace{4.8cm}
      \times \big( \sum_{k_3,\dots,k_{L+3}}
      [k_3 k_4] \;\eps^{l_1 l_2 k_3 \cdots k_{L+3}} \;
      \eta_{k_5a} \cdots \eta_{k_{L+3} a}\big)
  \Big\} ~~~~~\\ \nonumber
  &=&
  \d^{(8)}(\text{ext})~
  D_3 \cdots D_{L+1}
  \Big\{
    \frac{1}{(2(L-1)!)^4}\;
    \prod_{a=1}^4
     \big( \sum_{k_3,\dots,k_{L+3}} [k_3 k_4]\;
     \eps^{l_1 l_2 k_3 \cdots k_{L+3}} \;
      \eta_{k_5a} \cdots \eta_{k_{L+3} a}\big)
  \Big\} \\[3mm]
  &=&
  [A B]^4 \; \d^{(8)}(\text{ext}) \, .
  \lab{Lloop}
\eea
$A$ and $B$ are the external legs on the subamplitude $I$, c.f.~figure
\ref{fig:ABCD}.

As a simple check that this result is correct, let the legs $A$ and $B$ be positive helicity gluons and take the two other external legs $C$ and $D$ to be negative helicity gluons. Then there is only one term in the spin sum, namely
\bea
  A_{L+3}(A^+,B^+,l_1^-,\dots,l_{L+1}^-)\;
  A_{L+3}(C^-,D^-,-l_{L+1}^+,\dots,-l_{1}^+) \, ,
\eea
whose ``spin sum factor'' is simply $\<C D\>^4 [A B]^4$.
This is exactly what our result \reef{Lloop} produces when the two 4th order derivative operators $D_C$ and $D_D$ of the external negative helicity gluons are applied.

Rewriting the prefactors to separate the dependence on the loop momenta, the full result for the $L$-loop $(L+1)$-line cut MHV generating function is then simply
\bea
  \lab{Lloop2}
  \boxed{
  ~ F_\text{anti-MHV $\times$ MHV}^{L\text{-loop MHV 4-point}}
   ~=~
   [AB]^4
   \Big(
     [AB] [B| l_1 |C\> \<C D\> \< D | l_{L+1} |A]
     \prod_{i=1}^{L} P_{i,i+1}^2
   \Big)^{-1} \d^{(8)} (\text{ext}) \, . \;}
\eea

The result \reef{Lloop2} of an $L$-loop calculation is strikingly simple, yet it counts the contributions of states of total $\eta$-count $0\le r \le 8$ distributed arbitrarily on the $L+1$ internal lines in Fig.~\ref{fig:ABCD}.

\subsubsection{1-loop triple cut spin sum with anti-MHV $\times$ MHV $\times$ MHV}

Consider the triple cut of a 1-loop amplitude. In section
\ref{s:triple} we evaluated a triple cut spin sum assuming that the amplitude was overall NMHV, such that the three subamplitudes were MHV. We now consider the case where the amplitude is overall MHV. The $\eta$-count then tells us that $r_I+r_J+r_K = 8+4 \times 3 = 20$. At least one of the subamplitudes has to be anti-MHV with $\eta$-count 4. Thus let us assume $I$ to be anti-MHV and $J$ and $K$ MHV. The result is non-vanishing only if $I$ is a 3-point amplitude, i.e.~it has only one external leg, which we will label $A$. This is illustrated in figure \ref{fig:triple3pt}.

\begin{figure}
\begin{center}
 \includegraphics[width=5.5cm]{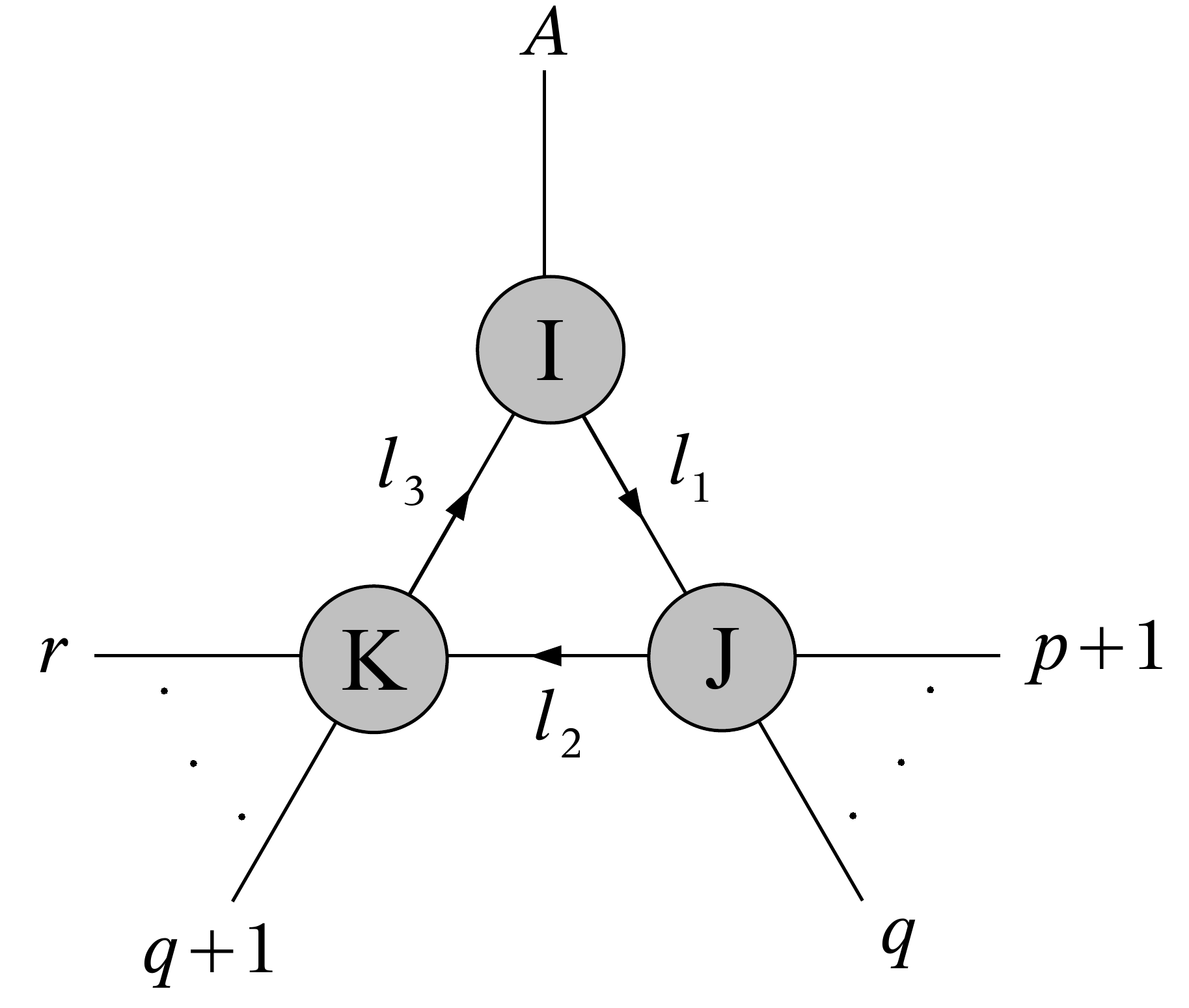}\\[5mm]
\end{center}
\caption{Triple cut of MHV 1-loop amplitude with anti-MHV subamplitude $I$ and MHV subamplitudes  $J$ and $K$.}
\lab{fig:triple3pt}
\end{figure}

We evaluate the spin sum using the anti-MHV and MHV generating functions. The expression for the spin sum factor requires some manipulation using momentum conservation, but the final result is simple:
\bea
  \boxed{
  ~F_\text{anti-MHV $\times$ MHV $\times$ MHV}^\text{1-loop MHV triple cut}
  ~=~
  \frac{\<r A \> \< A,p+1\> \< q,q+1\>~ [ l_1 A]^4 \< l_1 l_2 \>^4}
  {[A l_1] [l_1 l_3] [l_3 A]
    \< l_1, p+1 \> \< q l_2\> \< l_2 l_1 \> \< r l_3 \> \< l_3 l_2 \> \< l_2 , q+1\>}
      F_\text{MHV}^\text{tree} \, .~}
    ~~~
    \lab{tripMHV}
\eea
One simple check of this result is to assign all external states to be gluons, with $A$ and $B$ (on, say, subamplitude $J$) having negative helicity and the rest positive. Then the spin sum only contains one term, which gives a spin factor $[l_1 l_3]^4 \<l_1 B\>^4 \<l_2 l_3\>^4$. This must be compared with the result of \reef{tripMHV} with $D_A D_B$ applied, giving a spin factor $[ l_1 A]^4 \< l_1 l_2 \>^4 \< A B \>^4$. Momentum conservation on the 3-point subamplitude $I$ gives
\bea
[ l_1 A]^4 \< l_1 l_2 \>^4 \< A B \>^4
=[ l_3 A]^4 \< l_3 l_2 \>^4 \< A B \>^4
=[ l_3 l_1]^4 \< l_3 l_2 \>^4 \< l_1 B \>^4\, ,
\eea
so the results agree.


\subsubsection{4-loop anti-NMHV $\times$ NMHV spin sum}

The 5-line cut of the 4-point  4-loop amplitude includes an N$^3$MHV $\times$ NMHV sector in its unitarity sum. We use notation as in figure \ref{fig:ABCD}.
The tree subamplitudes are in this case 7-point functions and N$^3$MHV is therefore the same as anti-NMHV. We evaluate the spin sum using the NMHV and anti-NMHV generating functions.

Consider the anti-NMHV$_7(I)\;\times$ NMHV$_7(J)$ sector of the 5-line cut of the 4-loop 4-point amplitude. The intermediate state sum is straightforward to evaluate using the anti-NMHV generating function in the form \reef{antiNMHV}, choosing the lines $l_1$ and $l_2$ as the special lines 1 and 2.
The result of the intermediate spin sum is then
\bea
 \boxed{ \;
 F_\text{anti-NMHV$_7\times$ NMHV$_7$}^\text{4-loop MHV 5-line cut}
 ~=~
 \d^{(8)}({\rm ext})\;
 \frac{ \sum_{I_L, J_L} \overline{W}_{I_L} \, W_{J_L} \, (\rom{s.s.f})_{I_L,J_L}}
 {\< l_1 l_2\>^4
  \big( \prod_{i \in I} [i,i+1] \big) \big( \prod_{j \in J} \<j,j+1\> \big)} \;}
\eea
where
\bea
  \begin{array}{rcl}
  (\rom{s.s.f})_{I_L,J_L}
  &=&
  \bigg\{
  \sum_{j=3}^5
    \Big[
      \d_{l_j \in J_L} \,  \<l_j \,P_{J_L}\> \<l_1 l_2\>
        + \d_{l_1 \in J_L} \,  \<l_1 \, P_{J_L}\> \<l_2 l_j\>
        + \d_{l_2 \in J_L} \,  \<l_2 \, P_{J_L}\> \<l_j l_1\>
    \Big] \\[3mm]
    &&~~~~~
    \times
    \Big[
       \d_{l_j \in I_L}\, [l_j \, P_{I_L}] [A\,B]
      + \d_{A \in I_L} \, [A \, P_{I_L}] [B\, l_j]
      + \d_{B \in I_L} \, [B \, P_{I_L}] [l_j \, A]
    \Big]
  \bigg\}^4
  \end{array}
  ~~
\eea
The sum $\sum_{I_L,J_L}$ is over all 13 anti-MHV and MHV vertex diagrams in the expansions of the subamplitudes $I$ and $J$.
We have checked numerically that the cut amplitude generating function is independent of the two reference spinors $|X_I\>$ and $|X_J]$ from  the CSW prescription of $|P_{I_L}]$ and $|P_{J_L}\>$.

The complete spin sum for
 this cut of the 
4-loop 4-point amplitude contains the four contributions listed in the table in figure \ref{fig:gencut}. The anti-MHV $\times$ MHV contribution is obtained as the $L=4$ case of \reef{Lloop2} and we have here presented the result for the anti-NMHV $\times$ NMHV spin sum. The MHV $\times$ anti-MHV and NMHV $\times$ anti-NMHV contributions are obtained directly from these results.

\subsubsection{Other cuts of the 4-loop 4-point amplitude}

The full 4-loop calculation requires the study of unitarity cuts in which a 6-point subamplitude appears with all 6 lines internal and cut.  The simplest case is that of a 4-point function, hence overall MHV,   which can be expressed as
the product  $(2\to 3)(3\to 3)(3\to 2)$  of  3 sub-amplitudes.
See Figure~\ref{fig:2332diagram}.
\begin{figure}
\begin{center}
 \includegraphics[width=7cm]{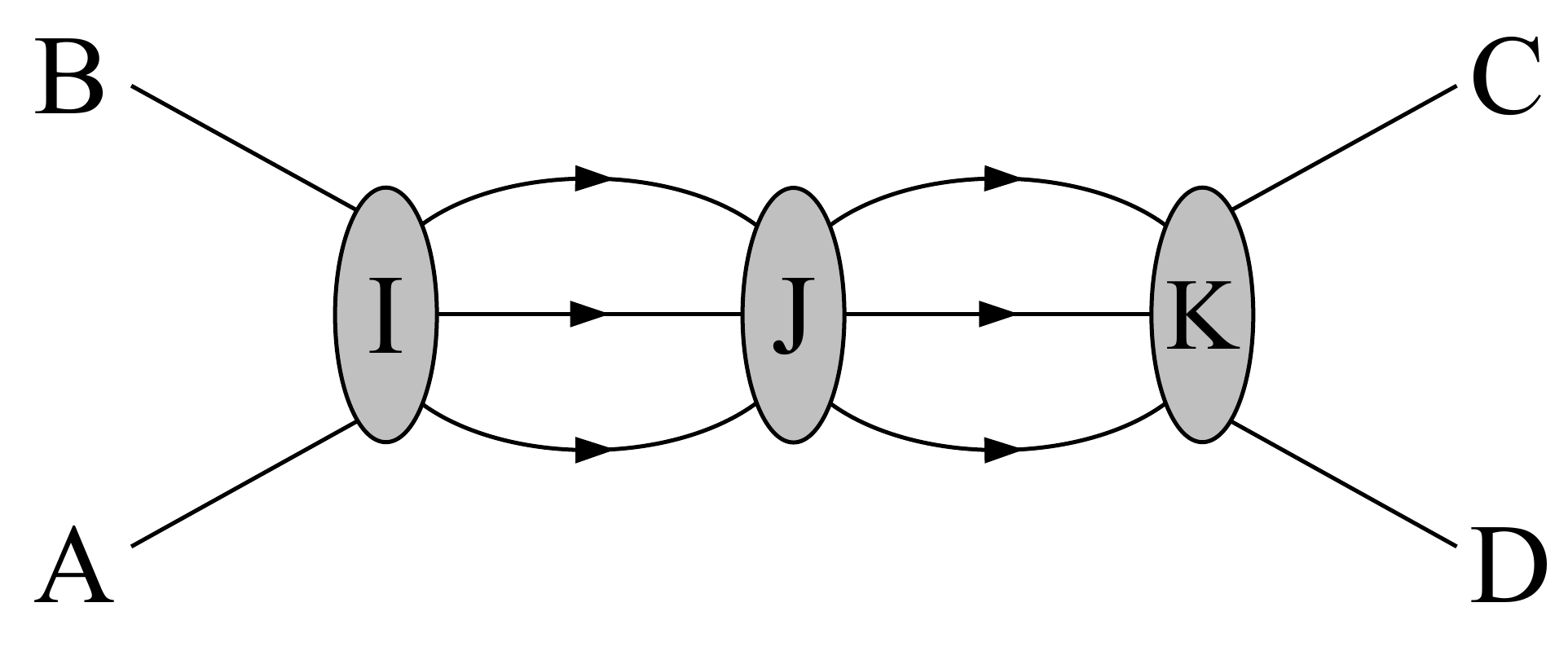}\\[3mm]
\end{center}
\vspace{0mm}
\caption{
A unitarity cut of  diagrams that contribute to the 4-point MHV amplitude at 4 loops. Note that subamplitude $J$ only connects to internal lines.}
\lab{fig:2332diagram}
\end{figure}
The spin sum requires
integration over the $6\times 4 = 24$ $\h_{ia}$ variables of the internal lines, and
 each term contains 8 of the 16 Grassmann variables $\h_{Aa}.\h_{Ba},\h_{Ca},\h_{Da}$ associated with the external states.
Thus, before any integrations, we are dealing with a product of generating functions containing monomials of degree $8+24 = 32$.   The full unitarity sum contains several sectors in which the 32 $\h$'s are split as
\bea \lab{sectors}
\begin{array}{rcl}
8+16+8 &
 ~~\leftrightarrow~~~ 
I \times J \times K  ~= & 
\rom{MHV}_5\times ~ \overline{\rom{MHV}}_6 ~\times \rom{MHV}_5 
\label{case1}\\[1mm]
12 + 8 + 12 &&  
\overline{\rom{MHV}}_5\times ~\rom{MHV}_6 ~\times \overline{\rom{MHV}}_5\\[1mm] 
8+ 12 +12 &&  
\rom{MHV}_5\times \rom{NMHV}_6\times \overline{\rom{MHV}}_5\label{case2}\\[1mm]
12+12+8 &&
\overline{\rom{MHV}}_5\times \rom{NMHV}_6\times  \rom{MHV}_5.
\end{array}
\eea

We have carried out each of these spin sums
 explicitly. 
The first two cases are related to each other by conjugation (including conjugation of the external states). The last two are related by interchanging $I$ and $K$ and relabeling the internal momenta accordingly.
The 6-point NMHV amplitude can also be regarded as anti-NMHV.
We have calculated the spin sums in both ways, using the NMHV and anti-NMHV generating function for $J$. Different diagrams contribute in these calculations, but numerically the results agree (and they are independent of the reference spinors).


\setcounter{equation}{0}
\section{Valid 2-line shifts for any $\cn=4$ SYM amplitude}
\lab{s:2line}
In this section we turn our attention to 2-line shifts which give  recursion relations of the BCFW type.
We examine the behavior of a general $\cn=4$ SYM tree level amplitude
\begin{equation}\label{Angeneral}
    A_n(1 \dots i\dots j\dots n)
\end{equation}
under a 2-line shift
of type $[i,j\>$, i.e.
\be\lab{2shiftij}
  |\tilde{i}] = |i] + z |j] \, ,~~~~ |\tilde{i}\> = |i\> \, , ~~~~~~~~~~~
  |\tilde{j}] = |j] \, ,~~~~|\tilde{j}\> = |j\> - z |i\> \,,
\ee
 with $i\neq j$.  
We will show that for \emph{any}  amplitude $A_n$ with $n>4$, we can find a valid shift $[i,j\>$ such that the amplitude vanishes at least as fast as $1/z$ for large $z$. This implies that there is a valid BCFW recursion relation for any tree amplitude in $\cn=4$ SYM.

The strategy of our proof is the following. In~\cite{ccg} it was shown that amplitudes $A_n$ vanish at large $z$ under a shift $[i^-,j\>$ if line $i$ is a negative helicity gluon. We extend this result using supersymmetric Ward identities and show that
amplitudes vanish at large $z$ under any shift $[i,j\>$ in which the $SU(4)$ indices carried by the particle on line $j$ are a subset of the $SU(4)$ indices of particle $i$. We then show that such a choice of lines $i$ and $j$ exists for all non-vanishing amplitudes $A_n$ with $n \ge4$, except for some pure scalar 4- and 6-point amplitudes. The 4-point amplitude is MHV hence determined by the SUSY Ward identities. We then analyze the scalar 6-point amplitudes explicitly and find that there
 exist 
valid shifts $[i,j\>$ under which they vanish at large $z$.

\subsection{Large $z$ behavior from Ward identities}
For  an $\cn=4$ $n$-point tree level amplitude $A_n$  of the form~(\ref{Angeneral})
it was shown in~\cite{ccg} that
\begin{equation}\label{claim1}
    A_n(1\dots i^-\dots j\dots n)\sim O(z^{-1})~~\text{ under a $[i^-,j\>$ shift if $i$ is a negative helicity gluon,
    line $j$ arbitrary\,.~~}
\end{equation}
Now consider any amplitude $A_n$ which has two lines $i$ and $j$ such that the $SU(4)$ indices of line $j$ are a subset of
the $SU(4)$ indices of line $i$.
We will prove that the amplitude vanishes at large $z$ under the
BCFW $[i,j\>$-shift given in (\ref{2shiftij}). Specifically we will show that
\begin{equation}\label{claim2}
    \boxed{
    \text{\begin{tabular}{c}
    ~$A_n(1\dots i\dots j\dots n)\sim O(z^{-1})$
     \,,~ or better, 
    under the shift $[i,j\>$ \\[2mm]
    if all $SU(4)$ indices of $j$ are also carried by $i$\,.\end{tabular}}}
\end{equation}
Let $r_i$ be the number of $SU(4)$ indices carried by line $i$.
We will show~(\ref{claim2}) by (finite, downward) induction on $r_i$.
For $r_i=4$
 particle $i$  
is a negative helicity gluon and the statement~(\ref{claim2}) reduces to~(\ref{claim1}) which was proven in~\cite{ccg}.
Assume now that~(\ref{claim2}) is true for all amplitudes with  $r_i=\bar r$ for some $1\leq\bar r\leq 4$.
Consider any amplitude $A_n$ which has $r_i=\bar r-1<4$ and in which the $SU(4)$ indices of particle $j$ are a subset of the indices of particle $i$. We write this amplitude as a correlation function
\begin{equation}
  \lab{oriA}
    A_n(1\dots i\dots j\dots n)=\<\co(1)\dots\co(i)\dots\co(j)\dots\co(n)\>\,.
\end{equation}
Pick an $SU(4)$ index $a$ which is
\emph{not} carried by line $i$. Such an index exists because $r_i<4$. There exists an operator $\co^a(i)$ such that
\begin{equation}
    [\tilde Q_a,\co^a(i)]=\<\epsilon i\>\,\co(i)\,.
    ~~~~~\text{(no sum)}
\end{equation}
By assumption, the $SU(4)$ index $a$ is also not carried by line $j$, so
$ [\tilde Q_a,\co(j)]=0$.
We can now write a Ward identity based on the index $a$ as follows:
\begin{equation}\label{SUSYtrafo}
\begin{split}
    0~=~&\bigl\<\bigl[\tilde Q_a\,,\,\co(1)\dots\co^a(i)\dots\co(j)\dots\co(n)\bigr]\,\bigr\>\\[1mm]
    ~=~&\<\epsilon i\>\bigl\<\co(1)\dots\co(i)\dots\co(j)\dots\co(n)\bigr\>
    \,+\,\bigl\<\bigl[\tilde Q_a\,,\,\co(1)\dots\bigr]\,\co^a(i)\dots\co(j)\dots\co(n)\bigr\>\\[1mm]
    &+\bigl\<\co(1)\dots\co^a(i)\,\bigl[\tilde Q_a\,,\,\dots\bigr]\,\co(j)\dots\co(n)\bigr\>
    \,+\,\bigl\<\co(1)\dots\co^a(i)\dots\co(j)\,\bigl[\tilde Q_a\,,\,\dots\co(n)\bigr]\bigr\>\,.
\end{split}
\end{equation}
Let us choose
 $|\epsilon\>\sim|\ell\>\neq |i\>,|j\>$. 
Then the first term on the right hand side is the original amplitude \reef{oriA}, multiplied
by a non-vanishing factor $\<\ell i\>$, which does not shift under the $[i,j\>$-shift.
The remaining three terms on the right hand side of~(\ref{SUSYtrafo}) all involve the operators $\co^a(i)$ and $\co(j)$. The number of $SU(4)$ indices carried by $\co^a(i)$ is $r_i+1=\bar r$, and therefore, by the inductive assumption, each of the remaining amplitudes fall off at least as fast as $1/z$ under the $[i,j\>$-shift. They are multiplied by angle brackets of the form $\<\ell k\>$ with $k\neq i,j$. These angle brackets do not shift. Thus the last three terms on the right side of~(\ref{SUSYtrafo}) go as $1/z$ or better for large $z$. We conclude that the amplitude $A_n(1\dots i\dots j\dots n)$ also vanishes
 at least 
as $1/z$ for large $z$ under the $[i,j\>$-shift. This completes the inductive step and proves~(\ref{claim2}).

Our result implies, in particular, that any shift $[i,j^+\>$ gives a $1/z$ falloff for any state $i$. This is because a positive helicity gluon $j^+$ carries no $SU(4)$ indices, and the empty set is a subset of any set.

\subsection{Existence of a valid 2-line shift for any amplitude}

We have proven the existence of a valid recursion relation for any amplitude which admits a shift of the type~(\ref{claim2}).
Let us examine for which amplitudes such a shift is possible. In other words, we study which amplitudes
contain two lines $i$ and $j$ such that the $SU(4)$ indices carried by line $j$ are a subset of the indices carried by line $i$. For $n$-point functions with $n\ge 4$ we find:
\begin{itemize}
  \item {\bf Any amplitude which contains one or more gluons admits a valid shift}\\
    If the amplitude contains a negative helicity gluon we pick this particle as line $i$. On the other hand, if the amplitude contains a positive helicity gluon we pick the positive helicity gluon as line $j$. Independent of the
    choice of particle for the other shifted line,~(\ref{claim2}) guarantees that the amplitude vanishes for large $z$ under the shift $[i,j\>$.
  \item {\bf Any amplitude with  one or more positive helicity gluinos admits a valid shift}\\
    We pick the positive helicity gluino as line $j$. Denote the $SU(4)$ index carried by this gluino by $a$. If no other line carries this index $a$, the amplitude vanishes.
    Thus in a non-vanishing amplitude there must be at least one other line $i\neq j$ which carries the index $a$. As line $j$ does not carry indices other than $a$, we can apply~(\ref{claim2}) and conclude that the amplitude falls off at least as $1/z$ under the shift $[i,j\>$.
  \item {\bf Any amplitude with  one or more negative helicity gluinos admits a valid shift}\\
    This proof is the $SU(4)$ conjugate version of the proof above.
    Now we pick the negative helicity gluino as line $i$ and denote the $SU(4)$ index which is \emph{not} carried by this gluino by $a$. If all other lines carry this index $a$, the amplitude vanishes. To see this, pick any other line $k\neq i$. The operator $\co^a(k)$ on this line carries the index $a$, so there exists an operator $\co(k)$ which satisfies
    \begin{equation}
    [Q^a,\co(k)]=[\epsilon k]\,\co^a(k)\,.
    \end{equation}
    Picking
     $|\epsilon]\sim |i]$
    we obtain
    \begin{equation}\label{SUSYtrafo2}
    0=\bigl\<\bigl[Q^a\,,\,\co(1)\dots\co(i)\dots\co(k)\dots\co(n)\bigr]\,\bigr\>
     ~\sim~ 
    [ik]\bigl\<\co(1)\dots\co(i)\dots\co^a(k)\dots\co(n)\bigr\> \, .
    \end{equation}
    As $i\neq k$, $[ik]$ is non-vanishing, and we conclude that the amplitude must vanish.
    
    Thus in a non-vanishing amplitude there must be at least one other line $j\neq i$ which does not carry the index $a$. As line $i$ carries all indices except for $a$, we can apply~(\ref{claim2}) and again conclude that
     the 
    amplitude falls off at least as $1/z$ under the $[i,j\>$-shift.
  \item
    {\bf Any pure scalar amplitude $A_n$ with $n>6$ admits a valid shift}\\
    In a pure scalar amplitude each particle carries two $SU(4)$ indices. There are ${4\choose 2}=6$ different combinations of indices possible, corresponding to the six distinct scalars of $\cn=4$ SYM. Thus any pure scalar amplitude $A_n$ with $n>6$ must have at least two lines $i$ and $j$ with the same particle and thus with coinciding $SU(4)$ indices.
    Using~(\ref{claim2}) we find that the amplitude vanishes for large $z$ under the $[i,j\>$-shift.
\end{itemize}
We are left to analyze pure scalar amplitudes with $n\le 6$. Amplitudes containing two identical scalars admit a valid shift by~(\ref{claim2}). Thus
we need only check amplitudes which involve distinct scalars:

\begin{itemize}
\item[-] $n=4$: There are two types of inequivalent 4-point amplitudes with four distinct scalars. The first type is the constant amplitude $\< A^{12} A^{23} A^{34} A^{41}\>=1$, and $SU(4)$ equivalent versions thereof.
The only contribution to this amplitude is from the 4-scalar interaction in the Lagrangian.
Clearly, it does not a have any good 2-line shifts, but since it is MHV it can be determined uniquely from SUSY Ward identities.
    The second type of amplitudes are $SU(4)$ equivalent versions of
    \begin{equation}
        \< A^{12} A^{34} A^{23} A^{41}\>
        =\frac{\<13\>\<24\>}{\<12\>\<34\>}\,.
    \end{equation}
    For example, this amplitude  vanishes at large $z$ under a $[1,3\>$-shift, and thus admits a valid recursion relation.
\item[-] $n=5$: By $SU(4)$ invariance, there are no non-vanishing 5-point functions with 5 distinct external scalars.
\item[-] $n=6$: We perform explicit checks of the pure scalar amplitudes $A_6$ in which the external particles are precisely the
six distinct scalars of the theory, i.e.~the amplitudes involving the particles $A^{12}$, $A^{13}$, $A^{14}$, $A^{23}$, $A^{24}$, and $A^{34}$.
We find that all possible permutations of the color ordering of the six scalars give amplitudes which fall off as $1/z$ under a shift $[i,i+3\>$ for some choice of line $i$. This is done by explicitly computing each amplitude using the NMHV generating function, whose validity was proven in section~\ref{s:valid3}, and then numerically testing the $[i,i+3\>$-shifts for different choices of line $i$.
\end{itemize}

\medskip

We conclude that for any $\cn=4$ SYM amplitude with $n>4$ there exists at least one choice of lines $i$ and $j$ such that
under a $[i,j\>$-shift
\begin{equation}\label{2lineresult}
    A_n(1\dots \tilde{i}\dots \tilde{j}\dots n) 
    \to 0\quad\text{ as $z\to\infty$}\,.
\end{equation}
The results also holds for $n=4$, with the exception of the 4-scalar amplitude mentioned above.

The input needed for our proof of \reef{2lineresult} was the result \cite{ccg} that a $[-,j\>$-shift gives a $1/z$-falloff (or better)
for any state $j$.
In $\cn=4$ SYM, the validity of a $[-,j\>$-shift can also be derived from the validity of shifts of type $[-,-\>$ using supersymmetric Ward identities. Thus we could have started with less: to derive~(\ref{2lineresult}) it is sufficient to know that amplitudes vanish at large $z$ under any $[-,-\>$-shift.


\setcounter{equation}{0}
\section{Summary and Discussion}

In this paper we have explored  the
validity and application of recursion relations for $n$-point amplitudes with general external states in
$\cn =4$ SYM theory. We now summarize our results, discuss some difficulties which limit their extension to
$\cn =8$ supergravity, and comment on some recent related papers.

1. We were especially concerned with recursion relations following from 3-line shifts because these give the most convenient representations for NMHV amplitudes, namely the MHV vertex expansion. We were motivated by the fact that these representations are useful in the study of multi-loop amplitudes in
$\cn =4$ SYM, and it is important \cite{bernpc} to know that they are valid.
The expansion can be derived using analyticity in the
complex variable $z$ of the 3-line shift \reef{3lineshift} if the shifted amplitude vanishes as $z \to \infty$. We proved that this condition holds \emph{if the
3 shifted lines carry at least one common $SU(4)$ index}.  $SU(4)$ invariance guarantees that  at least one such shift such exists for any NMHV amplitude.
For shifts with no common index, there are examples of amplitudes which do not vanish at large $z$ and other examples which do. So the common index criterion is sufficient but not always necessary.

2. We reviewed the structure of the MHV vertex expansion in
order to emphasize properties which are important for our applications.  A valid 3-line shift, which always exists, is needed to derive the expansion but there is no trace of that shift in the final form of the expansion. For most amplitudes there are several valid shifts, and each leads to the same expansion, which is therefore unique.
The main reason for this is that the MHV subdiagrams depend only on holomorphic spinors $|i\>$ and $|P_I\>$
of the external and internal lines of a diagram.  These are
not shifted, since the shift affects only the anti-holomorphic spinors $|m_i]$ of the 3 shifted external lines, $m_1,m_2,m_3.$ These desireable properties allow the definition of a universal NMHV generating function which describes all possible $n$-point processes. This generating function is written as a sum of an ``over-complete'' set of diagrams which can potentially contribute. Particular amplitudes are obtained by applying a 12th order differential operator
in the Grassmann variables of the generating function, and each diagram then appears multiplied by its spin factor. The spin
factor vanishes for diagrams which do not contribute to
the MHV vertex expansion of a given amplitude. What remains are the diagrams,
each in correct form, which actually contribute to the expansion.

3. In \cite{BEF} it was shown how to use the MHV generating function to carry out the intermediate spin sums in the
unitarity cuts from which loop amplitudes are constructed from products of trees. In this paper we have used Grassmann integration to simplify and generalize the previously treated MHV level sums, and
we have computed several new examples of sums which require the NMHV
generating function on one or both sides of the unitarity cut.  The external
states in the cut amplitudes are arbitrary and we were able to describe this
state dependence with new generating functions.

4. It is well known that the full set of amplitudes in $\cn=4$ SYM theory includes
the anti-MHV sector. This contains the $n$-gluon amplitude in helicity configuration $A_n(++--\ldots--)$ and all others related by SUSY transformations. Each anti-MHV amplitude is the complex conjugate of an MHV amplitude, but this description is
not well suited to the evaluation of unitarity sums.  Similar remarks apply to
anti-NMHV amplitudes which include $A_n(+++--\dots-)$ and others related by
SUSY. For this
reason we developed generating functions for anti-MHV and anti-NMHV $n$-point amplitudes. We used a systematic prescription to convert any generating function to the conjugate generating function by conjugation of brackets and a simple transformation to a new function of the same Grassmann variables $\h_{ia}$.  We then performed 3- and 4-loop unitarity sums in which  anti-MHV or anti-NMHV amplitudes appear on one side of the cut and MHV or NMHV on the other side.

5. Our study of the large $z$ behavior of NMHV amplitudes required starting with a concrete representation for them on which
we could then perform a 3-particle shift. We used the BCFW
recursion relation which is based on a 2-line shift. It was very recently shown in \cite{ccg} that such a recursion relation
is valid for any amplitude in $\cn =4$ SYM which contains at
least one negative helicity gluon. Using SUSY Ward identities we
were able to remove this restriction. The BCFW recursion relation is valid for all amplitudes.\footnote{Except for one particular 4-scalar amplitude which is constant and thus inert under all shifts.}


\vspace{3mm}
It is natural to ask whether the properties found for recursion relations and generating functions in $\cn =4$ SYM theory are true in $\cn =8$ supergravity. Unfortunately the answer is that not all features carry over at the NMHV level.  One complication is that the shifted MHV subamplitudes which appear in the MHV vertex expansion involve the shifted spinors $|m_i],$ so the expansion is no longer shift
independent or unique.
Valid expansions can be established for many 6-point NMHV amplitudes, but
it is known \cite{BEF} that there are some
amplitudes which do not vanish at large $z$ for any 3-line shift.
In these cases one must fix the reference spinor $|X]$ such that the $O(1)$ term at $z \to \infty$ vanishes in order to obtain a valid MHV vertex expansion.

Concerning the 2-line shift recursion relations, there are amplitudes in $\cn = 8$ SG which do not admit any valid 2-line shifts. One example is the 6-scalar amplitude
$\big\<
    \phi^{1234} \phi^{1358}
    \phi^{1278} \phi^{5678}
    \phi^{2467} \phi^{3456}
  \big\>$. No choice of two lines satisfies the index subset criteria needed in section \ref{s:2line} above, and a numerical analysis shows that there no valid 2-line shifts \cite{BEF}, contrary to the analogous $\cn=4$ SYM cases.


\vspace{3mm}
We  would like to mention several very recent developments which provide, in effect, new versions of generating functions for amplitudes in $\cn = 4$ SYM theory.

The paper \cite{dhks} presents  expressions for tree and loop amplitudes based on the dual conformal symmetry \cite{dks,alday}.  This symmetry can be proven at
tree level using an interesting new recursion relation \cite{qmc} for amplitudes with general external states.
The formula for NMHV tree amplitudes in \cite{dhks} has the feature that it does not contain the arbitrary reference spinor that characterizes the MHV vertex expansions of \cite{csw}.  Dual conformal symmetry appears to be a fundamental and important property of on-shell amplitudes, but the presence of a reference spinor may well be an advantage. Indeed,  MHV
vertex expansions provide expressions for amplitudes that are quite easy to
implement in numerical programs, and the test that the full amplitudes are independent of the reference spinor is  extremely useful in practical applications.

The paper \cite{sok2} has several similarities with our work. They use the same SUSY generators devised in
\cite{BEF} and used here, the MHV generating function of \cite{nair} is common, and for $n=3,4,5$ the anti-MHV generating functions coincide. For $n \ge 6$ there are apparent differences
in the representation of NMHV amplitudes, since the MHV vertex
expansion is not directly used in \cite{dhks,sok2} and there is no reference spinor. It could be instructive to explore the
relation between these representations. In \cite{sok2} the
application of generating functions to double and triple cuts of 1-loop amplitudes initiated in \cite{BEF} and studied above are extended to quadruple cuts with interesting results for the box coefficients which occur.

The very new paper \cite{nima2} uses the fermion coherent state formalism to derive a new type of tree-level recursion relation for the entire set of $\cn=4$ amplitudes. There are many other intriguing ideas to study here.

\section*{Acknowledgements}
We are grateful to Zvi Bern and David Kosower for very valuable discussions and suggestions leading to this work. We thank Clifford Cheung for sharing with us his results on the 2-line shifts. We have also benefitted from discussions with Lance Dixon, Iosif Bena, Paolo Benincasa,  John Joseph Carrasco and Pierre Vanhove.

HE and DZF thank the organizers of the workshop ``Wonders of Gauge theory and Supergravity'' in Paris June 23-28, 2008, where this work was initiated. HE thanks Saclay and LPT-ENS for their hospitality.
HE and DZF gratefully acknowledge funding from the MIT-France exchange program.
MK would like to thank the members of Tokyo University at IPMU and Komaba campus for stimulating discussions and hospitality in the final stages of this work.

HE is supported by a Pappalardo Fellowship in Physics at MIT.
DZF is supported by NSF grant PHY-0600465. The work of MK was partially supported by the World Premier International
 Research Center Initiative of MEXT, Japan.
All three authors are supported by the US  Department of Energy through cooperative research agreement DE-FG0205ER41360.


\appendix

\setcounter{equation}{0}
\section{$3$-line shifts of NMHV amplitudes with a negative helicity gluon}\label{app17}
In section \ref{s:valid3} we considered NMHV $n$-point amplitudes $A_n(1^-,\ldots,m_2,\ldots, m_3,\ldots,n)$,
with particle 1 a negative helicity gluon and $m_2$ and $m_3$ sharing at least one common $SU(4)$ index. We claimed that
one always obtains a valid MHV vertex expansion from the $3$-line shift $[1,m_2,m_3|$.
In this appendix we provide the detailed proof of this claim.
As a starting point we use the result~\cite{ccg} that a $[1^-,\ell\>$-shift of any tree amplitude of $\cn = 4$ SYM falls off at least as $1/z$ for large $z$, for any choice of particle $\ell\neq1$.
The $[1^-,\ell\>$-shift therefore gives a valid recursion relation without contributions from infinity.

Our strategy is then as follows. In
section~\ref{1ell}, we first express the NMHV amplitude $A_n$ in terms of the recursion relation following a $[1^-,\ell\>$ shift and examine the resulting diagrams. In section \ref{3shift}, we perform a secondary $[1,m_2,m_3|$ shift on the vertex expansion resulting from the $[1^-,\ell\>$ shift. We pick particle  $\ell$ for the first shift such that it is non-adjacent to lines $m_2$ and $m_3$. This is always possible for $n \ge 7$ (except for one special case at $n=7$ which we examine separately in section \ref{n7}). We show that for large $z$ each diagram in the $[1^-,\ell\>$-expansion falls off at least as $1/z$ under the $[1,m_2,m_3|$-shift, \emph{provided} all NMHV amplitudes $A_{n-1}$ fall off as $1/z$ under a 3-line shift of this same type.
This allows us to prove the falloff under the shift inductively in section \ref{secinduction}.
In section \ref{n6} we explicitly verify the falloff for $n=6$ which validates the induction and completes the proof.


\subsection{Kinematics and diagrams of the $[1^-,\ell\>$ shift}
\lab{1ell}

The $[1^-,\ell\>$-shift is defined as
\bea
  |\tilde{1}] = |1] + z |\ell] \, ,~~~~ |\tilde{1}\> = |1\> \, , ~~~~~~~~~~~
  |\tilde{\ell}] = |\ell] \, ,~~~~|\tilde{\ell}\> = |\ell\> - z |1\> \,,
\eea
where particle $1$ is a negative helicity gluon, while line $\ell$ is arbitrary.
Consider a diagram of the $[1^-,\ell\>$-expansion with internal momentum $\tilde{P}_{1K} = \tilde{1} + K$.
The condition that the internal momentum is on-shell fixes the value of $z$ at the pole to be
$z_{1K}=\frac{P_{1K}^2}{\<1|K|\ell]}$, so that the shifted
 spinors 
at the pole are
\bea
  |\tilde{1}] ~=~ |1] + \frac{P_{1K}^2}{\<1|K|\ell]} |\ell] \, ,
  \hspace{1cm}
  |\tilde{\ell}\> ~=~ |\ell\> -  \frac{P_{1K}^2}{\<1|K|\ell]} |1\> \, .
\eea
At the pole, the internal momentum $\tilde{P}_{1K}$ can be written as
\bea
  \lab{P1Kc}
  (\tilde{P}_{1K})^{\da\b}=\frac{P_{1K}|\ell]\,\<1|P_{K}}{\<1|K|\ell]}\,.
\eea
This expression factorizes because $\tilde{P}_{1K}$ is null. It is then convenient to define spinors associated with $\tilde{P}_{1K}$ as
\bea
  |\tilde{P}_{1K}\> ~=~  \frac{P_{1K}|\ell]\,\< 1\ell \>}{\<1|K|\ell]}\, ,
   \hspace{1cm}
  [ \tilde{P}_{1K} | ~=~ \frac{\<1|P_{K}}{\< 1\ell \>} \, .
\eea
For future reference, we also record a selection of spinor products:
\bea
 \lab{id2}
 &&
 \< \tilde{1} \tilde{\ell} \> = \<1 \ell \>\, ,~~~~~~~
 \< \tilde{1} \tilde{P}_{1K} \> = \< 1 \ell \> \, ,~~~~~~~~
 \< \tilde{\ell} \tilde{P}_{1K} \>
 = - \frac{\< 1 \ell \>\,  P_{1K \ell}^2}{\< 1| K | \ell]}\, ,\\[2mm]
 \lab{id3}
 &&
 [\tilde{1} \tilde{\ell}] = [1 \ell]\, ,~~~~~~~~~
 [\tilde{1} \tilde{P}_{1K}] = - \frac{K^2}{\< 1 \ell \>} \, ,~~~~~~
 [\tilde{\ell} \tilde{P}_{1K}] = - \frac{\< 1 | K | \ell ]}{\< 1 \ell \>} \, .
\eea

\medskip

We write the diagrams resulting from the $[1^-,\ell\>$ shift such that line 1 is always on the Left sub-amplitude L and line $\ell$ on the Right sub-amplitude R.
We denote the total number of legs on the L (R) subamplitude by $n_L$ ($n_R$). Applied to the $n$-point amplitude $A_n$, we have $n_L + n_R = n+2$.

We can use kinematics to rule out the following classes of L $\times$ R diagrams:
\begin{itemize}
\item There are \emph{no} MHV $\times$ MHV diagrams with $n_R=3$.

Proof: On the R  side we would have a $3$-vertex with lines $\ell$, $P_{1K}$ and one more line $y\in\{\ell-1,\ell+1\}$.
The R vertex is MHV when $r_\ell+r_y + r_{P} = 8$, which requires $r_\ell+r_y \ge 4$. The value of the R subamplitude is fixed by ``conformal symmetry'' (see sec 5 of \cite{BEF})
\bea
  A_R =
  \< y \, \tilde{\ell}\>^{r_y+r_\ell - 5}
  \< y \, \tilde{P}_{1K}\>^{3-r_\ell}
  \< \tilde{\ell} \, \tilde{P}_{1K}\>^{3-r_y} \, .
\eea
Upon imposing momentum conservation $P_{1K}=-p_\ell-p_y$, short calculations yield
\bea
  \< y \, \tilde{\ell}\>
  =  \< y \, \tilde{P}_{1K}\>
  = \< \tilde{\ell} \, \tilde{P}_{1K}\>
  = 0 \, .
\eea
So all three angle brackets entering $A_R$ vanish. Since $A_R$ has one more angle bracket in the numerator than in the denominator, the amplitude vanishes in the limit where we impose momentum conservation.

\item There are \emph{no} anti-MHV $\times$ NMHV diagrams.

Proof: On the L  side we would have a $3$-vertex with lines $1$, $-P_{1K}$ and one more line $x\in\{2,n\}$.
For this vertex to be anti-MHV we need $r_1 + r_x + r_P= 4$, and since $r_1=4$, this diagram only exists if line $x$ is a positive  helicity gluon, i.e. $r_x=0$. The value of this subamplitude is
\bea
  A_L = \frac{[x \, \tilde{P}_{1x}]^3}{[\tilde{1} x][\tilde{1}\tilde{P}_{1x}]}\, ,
\eea
but using momentum conservation we find that each square bracket vanishes:
\bea
   [x\, \tilde{P}_{1x}]= [\tilde{1} x]=[\tilde{1} \tilde{P}_{1x}]  = 0 \, .
\eea
As $A_L$ has more square brackets in the numerator than in the denominator we conclude that the L subamplitude vanishes.

\end{itemize}

\noindent Thus only the following two types of diagrams contribute to the recursion relation:

\begin{description}
\item[Type A:] MHV $\times$ MHV diagrams with $n_L \ge 3$ and $n_R \ge 4$.
\item[Type B:] NMHV $\times$ anti-MHV diagrams with $n_L =n-1$ and $n_R =3$.
\end{description}

We have thus obtained a convenient representation of the amplitude $A_n(1^-,\ldots,m_2,\ldots, m_3,\ldots,n)$ using the
2-line shift $[1^-,\ell\>$.  We will now use this representation of the amplitude to examine its behavior under a 3-line shift.


\subsection{The secondary $|1,m_2,m_3]$ shift}
\lab{3shift}
We now
act with the 3-line shift $|1,m_2,m_3]$ whose validity we want to prove.
The shift $[1,m_2,m_3|$ is defined as
\bea\label{3lineapp17}
  |\hat{1}] &=& |1] + z \<m_2 m_3 \> |X]\, , \nonumber\\
  |\hat{m}_2] &=& |m_2] + z \<m_3 1\> |X]\, , \\
  |\hat{m}_3] &=& |m_3] + z \<1 m_2 \> |X]\, .\nonumber
\eea
By assumption, the lines $m_2$ and $m_3$ have at least one $SU(4)$ index in common. Such a choice is possible for any NMHV amplitude. Up to now, we have not constrained our choice of line $\ell$ for the primary shift. It is now convenient to choose an $\ell\notin \{m_{2},m_3\}$ \emph{which is not adjacent} to either $m_2$ or $m_3$.
This is always possible for $n\geq7$, except for one special case with $n=7$ which we examine separately below.
We will now show that under the shift~(\ref{3lineapp17}), amplitudes vanish at least as $1/z$ for large $z$, \emph{provided} this falloff holds for all NMHV amplitudes with $n-1$ external legs under the same type of shift. This will be the inductive step of our proof.

The action of the shift on the recursion diagrams depends on how $m_2$ and $m_3$ are distributed between the L and R subamplitudes. We need to consider three cases: $m_2, m_3 \in L$, $m_2, m_3 \in R$ and
 $m_2 \in R, m_3 \in L$ (or, equivalently, $m_2 \in L, m_3 \in R$).

\subsubsection*{Case I: $m_2, m_3 \in R$}
The legs on the R subamplitude include $\ell, m_2, m_3, \tilde{P}_{1K}$ as well as at least one line separating $\ell$ from $m_{2,3}$, so $n_R \ge 5$. Hence the diagram must be of type A: MHV $\times$ MHV.

Since $m_2, m_3 \notin K$, the angle-square bracket $\<1|K|\ell]$ is unshifted, but
\bea
  \hat{P}_{1K}^2 = P_{1K}^2 - z \< m_2 m_3 \> \<1|K|X] \, ,
\eea
and therefore
\bea\label{Case1shifts}
  |\hat{\tilde{1}}] &=& |\tilde{1}]
  + z \<m_2 m_3\>\frac{\<1|\ell|X]}{\<1|K|\ell]} |\tilde{P}_{1K}] \, ,\nonumber\\
  |\hat{\tilde{P}}_{1K}\> &=& |\tilde{P}_{1K}\>
  - z \<m_2 m_3\>\frac{\<1|\ell|X]}{\<1|K|\ell]} |\tilde{1}\> \, ,\\
  |\hat{\tilde{\ell}} \> &=& |\tilde{\ell} \>
  + z  \<m_2 m_3\>\frac{\<1| K |X]}{\<1|K|\ell]} |\tilde{1}\> \,\nonumber .
\eea
while $|\hat{\tilde{P}}_{1K}]=|\tilde{P}_{1K}]$.
For arbitrary external lines $a\notin\{\hat{\tilde{\ell}},\tilde{1}\}$ one can check that
\bea\label{Case1ab1}
    \< a \hat{\tilde{P}}_{1K}\>~\sim~O(z)\,,\qquad \< a \hat{\tilde{\ell}}\>~\sim~O(z)\,.
\eea
The remaining angle brackets shift as follows:
\bea\label{Case1ab2}
    \< \hat{\tilde{\ell}} \hat{\tilde{P}}_{1K}\>~\sim~O(z)\,,\qquad
    \< \tilde{1} \hat{\tilde{P}}_{1K}\> ~\sim~ O(1)\,,\qquad
    \< \tilde{1} \hat{\tilde{\ell}}\>~\sim~O(1)\,,
\eea
while all other angle brackets are $O(1)$.

We can now examine the effect of the $|1,m_1,m_2]$ shift on the MHV $\times$ MHV diagram:
\begin{itemize}
\item \underline{$A_L$}:
On the $L$ subamplitude, only $|\hat{\tilde{1}}]$ and $|\hat{\tilde{P}}_{1K}\>$ shift. The shift is a (rescaled) $[\tilde{1}^-,\tilde{P}_{1K}\>$-shift and thus $A_L$ falls off at least as $1/z$ for large $z$ by the results of~\cite{ccg}.
\item The propagator gives a factor $1/z$.
\item \underline{$A_R$}:
Since line $\tilde{1}$ belongs to the L subamplitude, $\< \tilde{1} \hat{\tilde{P}}_{1K}\>$ and $\< \tilde{1} \hat{\tilde{\ell}}\>$ do not appear in $A_R$ and it thus follows from~(\ref{Case1ab1}) and~(\ref{Case1ab2}) that all angle brackets in $A_R$ which involve $ \hat{\tilde{P}}_{1K}$ or $\hat{\tilde{\ell}}$ are $O(z)$ under the shift. The numerator of $A_R$ consists of four angle brackets and grows at worst as $z^4$ for large $z$.
If $\tilde{P}_{1K}$ and $\tilde{\ell}$ are consecutive lines in the R subamplitude, then the denominator of $A_R$ contains three shifted angle bracket and therefore goes as $z^3$. Otherwise, the denominator contains four shifted angle brackets
and goes as $z^4$. Thus the worst possible behavior of $A_R$ is $O(z)$.
\end{itemize}
We conclude that any diagrams with $m_2, m_3 \in R$ fall off as $O(z^{-1})\,\frac{1}{z}\,O(z^1) \sim O(z^{-1})$ for large $z$.


\subsubsection*{Case II:\footnote{Note that the case of $m_2 \in L, m_3 \in R$ is obtained from this one by taking $m_2 \lra m_3$ and $z \lra -z$ in all expressions.} $m_3 \in L, m_2 \in R$.}

Since we chose $\ell$ non-adjacent to $m_2$, the R subamplitude must have $n_R \ge 4$ legs. Hence all diagrams in this class must be of type A (MHV $\times$ MHV).

We need to analyze the large $z$ behavior of the angle-brackets relevant for the MHV subamplitudes.
As $z \to \infty$ we find that the leading behavior of $|\hat{\tilde{\ell}}\>$ and $|\hat{\tilde{P}}_{1K}\>$ is given by
\bea
  |\hat{\tilde{\ell}}\> & = & |\ell \>
  - \frac{\<m_2|1+K|X]}{\<1 m_2 \> [\ell X]} |1\> + O(z^{-1}) \, ,\\[2mm]
  \lab{IIPang}
  |\hat{\tilde{P}}_{1K}\> &=&
  \frac{\< 1 \ell \>}{\< 1 m_2 \>} |m_2 \> + O(z^{-1}) \, .
\eea
Short calculations then yield the following large $z$ behavior for the relevant angle brackets:
\bea\label{Case2ab1}
    \< m_2 \,\hat{\tilde{\ell}} \>~\sim~O(1)\,,\qquad
    \< \hat{\tilde{\ell}} \,\hat{\tilde{P}}_{1K} \>~\sim~O(1)\,,\qquad
    \< a \,\hat{\tilde{P}}_{1K} \> ~\sim~ O(1)\qquad\text{for any }a\notin\{m_2,\hat{\tilde{\ell}}\}\,,
\eea
but
\bea\label{Case2ab2}
    \< m_2\, \hat{\tilde{P}}_{1K} \> ~\sim~ O(z^{-1})\,.
\eea
To derive these falloffs, we used
\begin{equation}
    \<m_2|1+K+\ell|X]\neq 0\,,
\end{equation}
which holds because the R subamplitude has more than $3$ legs, as noted above.

Now consider the effect of the secondary shift on the MHV $\times$ MHV diagram:
\begin{itemize}
\item \underline{$A_L$}: It follows from~(\ref{Case2ab1}) that all angle brackets in the L subamplitude are $O(1)$, so $A_L \sim O(1)$.
\item The propagator gives a factor of $1/z$.
\item \underline{$A_R$}: All angle brackets are $O(1)$, except for $\< m_2 \hat{\tilde{P}}_{1K} \>$ which is $O(z^{-1})$ according to~(\ref{Case2ab2}). Note that on the L MHV subamplitude, the internal line $\tilde{P}_{1K}$ \emph{cannot have the common $SU(4)$ index of $m_2$ and $m_3$}, because this index is already carried by lines 1 and $m_3$. Therefore, $\tilde{P}_{1K}$ on the R subamplitude
     \emph{must have}
    this index in common with $m_2$. The ``spin factor''  in the numerator of the MHV subamplitude $A_R$ thus includes at least one factor of $\<m_2 \tilde{P}_{1K} \>$. If lines $m_2$ and $\tilde{P}_{1K}$ are non-adjacent in the $R$ subamplitude then all angle brackets in the denominator are $O(1)$ according to~(\ref{Case2ab1}). On the other hand, if lines $m_2$ and $\tilde{P}_{1K}$ are adjacent the denominator of $A_R$ also contains one factor of $\<m_2 \tilde{P}_{1K} \>$ and is thus $O(z^{-1})$. We conclude that at worst $A_R \sim O(1)$. Note that the common index of lines $m_2$ and $m_3$ was crucial to draw this conclusion.
\end{itemize}
We conclude that any diagrams with $m_3 \in L, m_2 \in R$ falls off at least as $O(1)\,\frac{1}{z}\,O(1) \sim O(z^{-1})$ for large $z$. The same argument holds also for the case $m_2 \in L, m_3 \in R$.


\subsubsection*{Case III: $m_2, m_3 \in L$}

 As the three lines  $1$, $m_2$ and $m_3$ all share a common $SU(4)$ index, the L subamplitude must be NMHV in order to be non-vanishing. Thus there can be no MHV $\times$ MHV diagrams in this class. All amplitudes must be of type B (NMHV $\times$ anti-MHV).
 The right subamplitude is anti-MHV and must have $n_R=3$ legs in order for the
diagram to be non-vanishing.
 The secondary shift acts on the L subamplitude as a 3-line shift $[\tilde{1},m_2,m_3|$.
As particle $1$ is a negative helicity gluon and as lines $m_2$ and $m_3$ share at least one common $SU(4)$ index,
 the shift is precisely of the same type as the original secondary 3-line shift.
 This shift acts \emph{only} on the L subamplitude, which has $n-1$ legs.
\begin{itemize}
\item \underline{$A_L$}: The L subamplitude $A_L$ goes as $1/z$ \emph{provided} a $[1,m_2,m_3|$-shift with line $1$ a negative helicity gluon and lines $m_2$ and $m_3$ sharing a common $SU(4)$ index is a good shift for
 amplitudes with $n-1$ legs.
\item The propagator is unshifted and thus $O(1)$.
\item \underline{$A_R$}: The right subamplitude is unshifted and thus $O(1)$.
\end{itemize}
We conclude that any diagrams with $m_2,m_3 \in L$ fall off as as $O(z^{-1})\,O(1)\,O(1) \sim O(z^{-1})$ for large $z$, assuming the validity of the same type of shift for $n-1$ legs.

\medskip
In summary, the diagrams resulting from the $|1^-\ell\>$ vertex expansion of the amplitude $A_n$ all fall off at least as fast as $1/z$ under the secondary shift $[1^-,m_2,m_3|$. For Case III, we needed to assume that a 3-line shift of type $[1^-,m_2,m_3|$
gives at least a  falloff of $1/z$ for $(n-1)$-point amplitudes of the same type. We can thus use a simple inductive argument to show the validity of the shift for all $n\geq8$. We will afterwards explicitly prove the falloff for $n=6,7$.

\subsection{Induction}\label{secinduction}
Let us assume that for some $n\geq7$, all
$\cn=4$ SYM NMHV $n$-point amplitudes $A_n$ satisfy
\bea \lab{assume}
  A_n(1,x_2,\dots,x_n) \to O(z^{-1})
\eea
under any $[1,m_2,m_3|$-shift with
 line $1$ a negative helicity gluon and lines
$m_{2,3}$ sharing at least one common $SU(4)$ index.
Consider now a $[1,m'_2,m'_3|$-shift on
$A_{n+1}(1,x'_2,\dots,x'_{n+1})$, again with $m'_{2,3}$ chosen to share at least one common $SU(4)$ index. We have shown above that $\ell$ can always be chosen such that all MHV $\times$ MHV vertex diagrams of the $[1^-,\ell\>$-shift  fall off at least as $1/z$ under the $[1,m'_2,m'_3|$-shift. Under the assumption \reef{assume}, we have shown that NMHV$_n$ $\times$ anti-MHV$_3$ vertex diagrams will also fall off at least as $1/z$.

\medskip

For the cases $n=6,7$, our inductive step is not applicable to all diagrams because we \emph{cannot} always pick line $\ell$ non-adjacent to $m_2$ and $m_3$. The diagrams where we cannot pick $\ell$ in this
way must be analyzed separately. For $n=7$, our reasoning above only fails for a small class of diagrams. Let us analyze this class of diagrams next.

\subsection{Special diagrams for $n=7$}
\lab{n7}

For 7-point amplitudes there is one color ordering of the three lines $1$, $m_2$ and $m_3$ which needs to be analyzed separately, namely
\bea
  \lab{n7amp}
  A_7(1,x_2,m_2,x_4,x_5,m_3,x_7) \, .
\eea
In this case we cannot choose $\ell$ to be non-adjacent to both $m_2$ and $m_3$. Instead choose $\ell = x_2$.
The analysis of all diagrams goes through as in section~\ref{3shift}, except
that Case II may now include a diagram of Type B (NMHV $\times$ anti-MHV), namely
\bea
  \lab{n7diag}
  A_L(\tilde{1}, -\tilde{P}_{1K},x_4,x_5,m_3,x_7) \frac{1}{P_{1K}^2}
  A_R(\tilde{\ell},m_2,\tilde{P}_{1K}) \, .
\eea
It appears because $\ell$ is adjacent to $m_2$.

As $z \to \infty$ we find
\bea
  |\hat{\tilde{1}}] & = &
     |\tilde{1}] +z \<m_2 m_3 \> |X]\, ,\\[2mm]
  |\hat{\tilde{P}}_{1K}] &=&
  |\tilde{P}_{1K}]- z \, \frac{\< 1 m_2 \> \< m_3 1 \>}{\< 1 \ell\>} |X] \,,\\
  |\hat{m}_3] &=& |m_3] + z \< 1 m_2\> |X]\,,
\eea
while $|\tilde{P}_{1K}\> $ remains unshifted.
Short calculations then yield the following large $z$ behavior for the relevant square brackets:
\bea\label{falloffsb}
[\hat{\tilde{1}} \hat{\tilde{P}}_{1K}]~\sim~ O(z)\,,\quad\qquad
 [a \hat{\tilde{P}}_{1K}]~\sim~ O(z)\,, \qquad
 [a \hat{\tilde{1}}]~\sim~ O(z)\qquad\text{for any }a\notin\{\hat{\tilde{1}},\hat{\tilde{P}}_{1K}\}\,.
\eea

Now consider the effect of the secondary shift on the NMHV $\times$ anti-MHV diagram:
\begin{itemize}
\item \underline{$A_L$}:
    After a rescaling
    $|\tilde{P}_{1K}\> \to  -\frac{\<1m_2\>}{\<1\ell\>} |\tilde{P}_{1K}\>$ and
    $|\tilde{P}_{1K}] \to  -\frac{\<1\ell\>}{\<1m_2\>} |\tilde{P}_{1K}]$, the shift acts exactly as a 3-line $[\tilde{1},\tilde{P}_{1K},m_3|$-shift. Note that line $\tilde{P}_{1K}$ on the $L$ side has at least one common index with $m_3$, because line $\tilde{P}_{1K}$ cannot carry this index on the $R$ side. In fact, this index is already carried by line $m_2$ in the right subamplitude, and as the right subamplitude is a $3$-point anti-MHV amplitude, each index must occur exactly once for a non-vanishing result.
    The behavior of the L subamplitude is thus given by the falloff of a $n=6$ amplitude under a $[\tilde{1},\tilde{P}_{1K},m_3|$ shift, in which line $1$ is a negative helicity gluon and lines $\tilde{P}_{1K}$ and $m_3$ share a common index.
\item The propagator gives a factor of $1/z$.
\item \underline{$A_R$}:
    The right subamplitude is a $3$-point anti-MHV and is thus the ratio of four square brackets in the numerator and three square brackets in the denominator. According to~(\ref{falloffsb}) all square brackets are $O(z)$ and we conclude that $A_R\sim O(z)$ for large $z$.
\end{itemize}
\emph{Provided} $n=6$ amplitudes fall off at least as $1/z$ under any
3-line shift $[1,m'_2,m'_3|$ in which $1$ is a negative helicity gluon and $m'_2$ and $m'_3$ share a common $SU(4)$ index, we conclude that the full amplitude \reef{n7amp}
goes as $O(z^{-1})\,O(z^{-1})\,O(z) \sim O(z^{-1})$ for large $z$.

\medskip

We have thus reduced the validity of our shift at $n=7$ to its validity at $n=6$. Let us now analyze $6$-point amplitudes.

\noindent

\subsection{Proof for $n=6$}
\lab{n6}
 Our analysis above for $n>6$ only used the fact that $\ell$ was non-adjacent to $m_{2,3}$ by ruling out certain diagrams of type B (NMHV $\times$ anti-MHV). For $n=6$, we cannot rule out these diagrams and will thus analyze them individually below.
Also, we will  estimate the large $z$ behavior of the NMHV$=$anti-MHV $5$-point subamplitudes that appear in the $[1^-,\ell\>$-shift expansion. This will complete the explicit proof of the desired large $z$ falloff at $n=6$, without relying on a further inductive step.

Choose a $[1, m_2, m_3|$-shift where $m_2$ and $m_3$ share a common $SU(4)$ index.
Using the freedom to reverse the ordering of the states
$123456 \to 165432$, there are six independent cases determined by the color ordering:

\begin{enumerate}
\item[(a)] $A_6(1,x_2,x_3,x_4,m_2,m_3)$
\item[(b)] $A_6(1,x_2,x_3,m_2,m_3,x_6)$
\item[(c)] $A_6(1,x_2,x_3,m_2,x_5,m_3)$
\item[(d)] $A_6(1,m_2,x_3,x_4,x_5,m_3)$
\item[(e)] $A_6(1,x_2,m_2,x_4,x_5,m_3)$
\item[(f)] $A_6(1,x_2,m_2,x_4,m_3,x_6)$
\end{enumerate}

\begin{itemize}
\item Consider first the four cases (a)--(d). In these amplitudes, $\ell$ can be chosen to be non-adjacent to $m_2$,$m_3$. We pick
    \bea
        \text{(a),(b),(c):}~~~\ell\to x_2\,,\quad
        \text{(d):}~~~\ell\to x_4\,.
    \eea
In all four situations the NMHV diagram is a Case III diagram ($m_2,m_3\in L$), so we have to
check the large $z$ behavior of the Left 5-point anti-MHV amplitude under the $[1,m_2,m_3|$-shift.
The denominator of the anti-MHV subamplitude will go as $z^4$ or $z^5$,
depending on whether the lines $1,m_2,m_3$ are consecutive or not. The numerator contains four square brackets,
at least one of which does not shift under $[1,m_2,m_3|$. This can be seen as follows. As the lines $1,m_2,m_3$ share
a common $SU(4)$ index, the other two lines in the $5$-point amplitude, $\tilde{P}_{1K}$ and (say) $y$ are the lines which do \emph{not} carry this index. Since the numerator of an anti-MHV amplitude contains precisely the square brackets of particles which do not carry a certain $SU(4)$ index, we conclude that there must be a factor of $[y \tilde{P}_{1K}]$ in the numerator.\footnote{To see this, one can also consider the conjugate MHV amplitude. Its numerator must contain a factor $\<y \tilde{P}_{1K}\>$ because the conjugated particles on lines $\tilde{P}_{1K}$ and $y$ share a common $SU(4)$ index. Conjugating back, we replace angle by square brackets and obtain the factor $[y \tilde{P}_{1K}]$ in the numerator. }
This factor does not shift under the $3$-line shift $[1,m_2,m_3|$, so the numerator grows as $z^3$ at worst.
The 5-point anti-MHV L-subamplitude will thus have at least a $1/z$ falloff.
As both the propagator and the right subamplitude remain unshifted, we conclude that the amplitudes (a)--(d) vanish at large $z$.

\item Next, consider the amplitude (e) above: $A_6(1,x_2,m_2,x_4,x_5,m_3)$. Choose
$\ell=x_4$. There are potentially two NMHV vertex diagrams: first, the
($\ell,x_5$) channel which has $m_2,m_3 \in L$ and we can thus apply the same argument we used for cases (a)--(d). The diagram of this channel therefore falls off at least as $1/z$.
Secondly, consider the ($m_2,\ell$)-channel.
This diagram has the same right subamplitude that we encountered for $n=7$ in the diagram~(\ref{n7diag}) above.
By the same analysis we conclude that $A_R \sim O(z)$.
 Note that the left subamplitude is given by
$A_L=A_5(\tilde{1},x_2,\tilde{P}_{1K},x_5,m_3)$.
 From eq.~(\ref{falloffsb}) we know that
all square brackets involving $\tilde{1},m_3,\tilde{P}_{1K}$ grow as $O(z)$, so
the numerator
will be at worst $O(z^4)$.
As the three shifted lines $\tilde{1},m_3,\tilde{P}_{1K}$ are not all consecutive,
  the denominator always goes as $z^5$. So $A_L \sim O(1/z)$, and as the propagator goes as $1/z$, we conclude
that the whole diagram is at worst $O(1/z)$.

\item Finally, consider case (f), $A_6(1,x_2,m_2,x_4,m_3,x_6)$. Choose $\ell=x_2$. Then the
NMHV vertex appears in the diagram with channel ($\ell,m_2$). This diagram can be treated just as the second diagram of case (e). To see this, note that $A_L=A_5(\tilde{1},\tilde{P}_{1K},x_4,m_3,x_6)$, so the three shifted lines, $\tilde{1},\tilde{P}_{1K},m_3$ are again not all consecutive. We conclude that this diagram also falls off at least as $1/z$ for large $z$.
\end{itemize}

\medskip
We conclude that for a NMHV $6$-point amplitude, any 3-line shift which involves a negative helicity gluon and two other states which share at least one common $SU(4)$-index, falls off at least as $1/z$ for large $z$. By the inductive argument of~\ref{secinduction} this immediately extends to all $n\geq6$ and completes the proof.


\setcounter{equation}{0}
\section{Anti-NMHV generating function from Anti-MHV vertex expansion}
\lab{appantiMHVVE}

In section \ref{s:Anmhv} we applied the Fourier transform prescription to obtain a generating function for anti-NMHV amplitudes,
\bea
  \lab{antinmhvGF}
  \bar{\cf}_{n} ~=~ \sum_{I}   \bar{\cf}_{n,I} \, ,
  \hspace{1.5cm}
  \bar{\cf}_{n,I}
  &=&
  \frac{1}{\prod_{i=1}^n [i,i+1]} \; \overline{W}_{I}\; 
   (\rom{S.F.})_I \, .
\eea
In this appendix we use $I$ to denote the diagrams of the anti-MHV vertex expansion.
The sum is over all diagrams in any anti-MHV vertex expansion and the spin factor is
\bea
  \lab{anSF}
  (\rom{S.F.})_I &=&
  \frac{1}{2^4}
  \prod_{a=1}^4 \sum_{i,j=1}^n  \sum_{k \in I_L}
  [i,j] [k P_{I_L}] \; \pa_{ia}\pa_{ja}\pa_{ka}\; \eta_{1a} \cdots \eta_{na} \, .
\eea

The purpose of this appendix is to use the anti-MHV generating function to  prove that \reef{antinmhvGF}-\reef{anSF} indeed is the correct generating function for an anti-MHV vertex diagram.

Consider any anti-MHV vertex diagram
\bea
  A^\text{anti-MHV}_{n_L} ( \dots ) \frac{1}{P_I^2}
  A^\text{anti-MHV}_{n_R} ( \dots ) \, .
\eea
The value of each anti-MHV subamplitude is found by applying the appropriate derivative operators to the anti-MHV generating function, whose correctness we have already confirmed in section \ref{s:Amhv}. The internal line must be an $SU(4)$ invariant, so its total order is 4. Given the external states, there is a unique choice of internal state, so the 4 internal line differentiations can be taken outside the product of anti-MHV generating functions. Thus the value\footnote{It was described in \cite{BEF} how to obtain the correct overall sign for the diagram.} of the diagram is
\bea
   D_{\rom{ext}} ~D_{I}~ \bar{F}_{n_L}(L)  \; \bar{F}_{n_R}(R)  \, .
\eea
This is true for any external states, hence the correct value of any anti-MHV vertex diagram is produced by the generating function
\bea
 \nonumber
 \bar{\cf}_{n,I}
 &=&D_{I}~ \bar{F}_{n_L}(L)  \; \bar{F}_{n_R}(R)
 ~=~ D_{I}~ \bigg(\frac{1}{\prod_{i \in L} [i,i+1]}\widetilde{\d^{(8)}(L)} ~~
 \frac{1}{\prod_{j \in R} [j,j+1]}\widetilde{\d^{(8)}(R)}  \bigg)\\
 &=& \frac{1}{\prod_{i=1}^n [i,i+1]} \; \overline{W}_{I}\; 
 \times (\text{S.F.})_I\, ,
 \lab{diagval}
\eea
with
\bea
  \lab{anSF2}
  (\rom{S.F.})_I &=&D_{I} \, \widetilde{\d^{(8)}(L)} ~  \widetilde{\d^{(8)}(R)} \, .
\eea
We have introduced
\bea
  \lab{antidelta}
  \widetilde{\d^{(8)}(L)}
  ~\equiv~\frac{1}{2^4} \prod_a
  \sum_{i, j \in L} [i j]\,\partial_{i}^a\partial_{j}^a
  \prod_{ k \in L}\eta_{k a} \, ,
\eea
and similarly for $\widetilde{\d^{(8)}(R)}$.

The prefactors of \reef{antinmhvGF} and \reef{diagval} are clearly the same, so we just need to prove that the spin factor in \reef{anSF2} is equal to that in \reef{anSF}. We start from \reef{anSF2} and write out the full expressions for the ``anti-delta-functions'' \reef{antidelta}, seperating out the internal line $I$ from the external lines,
\bea
\nonumber
(\rom{S.F.})_I
&=& D_I
 \bigg\{ \prod_a\Bigl[\sum_{i_L< j_L} [i_L j_L]\,\partial_{i_L}^a\partial_{j_L}^a
+ \sum_{ i_L} [i_L P_I]\,\partial_{i_L}^a \partial_{I}^a\Bigr]
\Bigl(\prod_{ k_L}\eta_{k_L a}\Bigr)\eta_{I a}\\
&&\qquad\qquad
~~~~\times\Bigl[\sum_{ i_R< j_R} [i_R j_R]\,\partial_{i_R}^a\partial_{j_R}^a
+ \sum_{ i_R} [i_R P_I]\,\partial_{i_R}^a \partial_{I}^a\Bigr]
 \Bigl(\eta_{I a}\prod_{ k_R}\eta_{k_R a}\Bigr)\bigg\}\, .
\eea
All lines $i_{L/R}, j_{L/R}, k_{L/R}$ are \emph{external} states on the L/R side of the vertex expansion. Evaluate first the derivatives $\partial_I$ inside the curly brackets to get
\bea
 \nonumber
(\rom{S.F.})_I
&=&D_I
\Bigl[\sum_{i_L< j_L} [i_L j_L]\,\partial_{i_L}^a\partial_{j_L}^a
\Bigl(\prod_{ k_L}\eta_{k_L a}\Bigr)\eta_{I a}
+(-)^{n_L-1} \sum_{ i_L} [i_L P_I]\,\partial_{i_L}^a\prod_{k_L}\eta_{k_L a}\Bigr]\\
&& ~~~~~~
\times\Bigl[\sum_{ i_R< j_R} [i_R j_R]\,\partial_{i_R}^a\partial_{j_R}^a\, \eta_{I a}\prod_{ k_R}\eta_{k_R a}
+ \sum_{ i_R} [i_R P_I]\,\partial_{i_R}^a\prod_{k_R}\eta_{k_L a}\Bigr]
\, .
\eea
Then perform the $D_I$ differentiation:
\bea
 \nonumber
(\rom{S.F.})_I
&=&\prod_a
\Bigl[(-)^{n_L-1}\sum_{i_L< j_L} [i_L j_L]\,\partial_{i_L}^a\partial_{j_L}^a \prod_{ k_L}\eta_{k_L a} \sum_{ i_R} [i_R P_I]\,\partial_{i_R}^a\prod_{k_R}\eta_{k_L a}\\
&&~~~~~~~~
+(-)^{2n_L-1} \sum_{ i_L} [i_L P_I]\,\partial_{i_L}^a\prod_{k_L}\eta_{k_L a}
\sum_{ i_R< j_R} [i_R j_R]\,\partial_{i_R}^a\partial_{j_R}^a\prod_{ k_R}\eta_{k_R a}
\Bigr] \, .
\eea
Now factor out the product of all $\eta$'s corresponding to the external lines to find
\bea
\nonumber
(\rom{S.F.})_I
&=&\frac{1}{2^4}\prod_a
\Bigl[\sum_{i_L, j_L,i_R} [i_L j_L][i_R P_I]\,\partial_{i_L}^a\partial_{j_L}^a \partial_{i_R}^a
- \sum_{ i_R, j_R,i_L}[i_R j_R][i_L P_I]\,
 \partial_{i_R}^a\partial_{j_R}^a\partial_{i_L}^a\Bigr]
 \prod_{\rom{ext}\; k}\eta_{k a}\,. \\[-4mm]
  \label{antiMHVvertexex}
\eea

Note that by the Schouten identity the antisymmetrized sum over 3 square brackets vanishes,
\begin{equation}\label{RRR}
\sum_{ i_R, j_R,k_R}[i_R j_R][k_R | \; \partial_{i_R}^a\partial_{j_R}^a\partial_{k_R}^a=0\,.
\end{equation}
We can thus remove the L restriction on the index $i_L$ in the second term in~(\ref{antiMHVvertexex})
and replace it by an index $m$ running over all external states. We then obtain
\begin{equation}
\begin{split}
&\sum_{ i_R, j_R,i_L}[i_R j_R][i_L P_I] \;
 \partial_{i_R}^a\partial_{j_R}^a\partial_{i_L}^a
=\sum_{ i_R, j_R,m}[i_R j_R][m P_I]\;
 \partial_{i_R}^a\partial_{j_R}^a\partial_{m}^a\\
&=-\sum_{ m,i_R, j_R}[m\, i_R][j_R P_I]\;
 \partial_{m}^a\partial_{i_R}^a\partial_{j_R}^a
-\sum_{ i_R,m, j_R}[j_R\, m ][i_R P_I]\;
 \partial_{j_R}^a \partial_{m}^a\partial_{i_R}^a\,.
\end{split}
\end{equation}
In the second line we have used the Schouten identity to split the sum.
This is done in order to complete the sum over $i_L$, $j_L$ in the first term of~(\ref{antiMHVvertexex}) to a sum over all external states $m$ and $n$
\bea
\nonumber
(\rom{S.F.})_I &=&
\frac{1}{2^4}\prod_a
\Bigl[\sum_{i_L, j_L,i_R} [i_L j_L][i_R P_I]\;
\partial_{i_L}^a\partial_{j_L}^a \partial_{i_R}^a\\
\nonumber
&&\hspace{1.5cm}
+\sum_{ m,i_R, j_R}[m i_R][j_R P_I]\;
 \partial_{m}^a\partial_{i_R}^a\partial_{j_R}^a
+\sum_{ i_R,m, j_R}[j_R m ][i_R P_I]\;
 \partial_{j_R}^a \partial_{m}^a\partial_{i_R}^a
\Bigr]\prod_{\rom{ext}\; k}\eta_{k a}\\[2mm]
&=&\frac{1}{2^4}\prod_a
\Bigl[\sum_{m, n,i_R} [m n][i_R P_I]\;
\partial_{m}^a\partial_{n}^a \partial_{i_R}^a\Bigr]\prod_{\rom{ext}\; k}\eta_{k a}\,.
\eea
We had to again use~(\ref{RRR}) in the last step. Finally the Schouten identity allows us to convert the sum over external momenta on the R to a sum over external momenta on the L subamplitude. The result is
\bea
(\rom{S.F.})_I &=&
\frac{1}{2^4}\prod_a
\Bigl[\sum_{m, n,i_L} [m n][i_L P_I]\;
\partial_{m}^a\partial_{n}^a \partial_{i_L}^a\Bigr]\prod_{\rom{ext}\; k}\eta_{k a}\,.
\eea
This is precisely the expected spin factor \reef{anSF} (given in the main text in \reef{barNFn0}) that our prescription predicts.



\begin{thebibliography}{99}

\bibitem{csw}
  F.~Cachazo, P.~Svrcek and E.~Witten,
  ``MHV vertices and tree amplitudes in gauge theory,''
  JHEP {\bf 0409}, 006 (2004)
  [arXiv:hep-th/0403047].


\bibitem{bcf}
  R.~Britto, F.~Cachazo and B.~Feng,
  ``New recursion relations for tree amplitudes of gluons,''
  Nucl.\ Phys.\  B {\bf 715}, 499 (2005)
  [arXiv:hep-th/0412308].

\bibitem{bcfw}
  R.~Britto, F.~Cachazo, B.~Feng and E.~Witten,
  ``Direct proof of tree-level recursion relation in Yang-Mills theory,''
  Phys.\ Rev.\ Lett.\  {\bf 94}, 181602 (2005)
  [arXiv:hep-th/0501052].


\bibitem{bdkreview}
  Z.~Bern, L.~J.~Dixon and D.~A.~Kosower,
  ``On-Shell Methods in Perturbative QCD,''
  Annals Phys.\  {\bf 322}, 1587 (2007)
  [arXiv:0704.2798 [hep-ph]].



\bibitem{nima1}
  N.~Arkani-Hamed and J.~Kaplan,
  ``On Tree Amplitudes in Gauge Theory and Gravity,''
  JHEP {\bf 0804}, 076 (2008)
  [arXiv:0801.2385 [hep-th]].

\bibitem{bdp}
  S.~J.~Bidder, D.~C.~Dunbar and W.~B.~Perkins,
  ``Supersymmetric Ward identities and NMHV amplitudes involving gluinos,''
  JHEP {\bf 0508}, 055 (2005)
  [arXiv:hep-th/0505249].

\bibitem{ccg}
  C.~Cheung,
  ``On-Shell Recursion Relations for Generic Theories,''
  arXiv:0808.0504 [hep-th].

\bibitem{risager}
  K.~Risager,
  ``A direct proof of the CSW rules,''
  JHEP {\bf 0512}, 003 (2005)
  [arXiv:hep-th/0508206].


\bibitem{nair}
  V.~P.~Nair,
  ``A Current Algebra For Some Gauge Theory Amplitudes,''
  Phys.\ Lett.\  B {\bf 214}, 215 (1988).



\bibitem{ggk}
  G.~Georgiou, E.~W.~N.~Glover and V.~V.~Khoze,
  ``Non-MHV tree amplitudes in gauge theory,''
  JHEP {\bf 0407}, 048 (2004)
  [arXiv:hep-th/0407027].

\bibitem{BEF}
  M.~Bianchi, H.~Elvang and D.~Z.~Freedman,
  ``Generating Tree Amplitudes in N=4 SYM and N = 8 SG,''
  arXiv:0805.0757 [hep-th].

\bibitem{efk2}
  M.~Kiermaier, H.~Elvang and D.~Z.~Freedman,
  ``Proof of the MHV vertex expansion for all tree amplitudes in N=4 SYM theory,''
  arXiv:0811.3624 [hep-th].

\bibitem{Drummond:2008cr}
 J.~M.~Drummond and J.~M.~Henn,
 ``All tree-level amplitudes in N=4 SYM,''
 arXiv:0808.2475 [hep-th].


\bibitem{pt}
  S.~J.~Parke and T.~R.~Taylor,
  ``An Amplitude for $n$ Gluon Scattering,''
  Phys.\ Rev.\ Lett.\  {\bf 56}, 2459 (1986).


\bibitem{bddk}
  Z.~Bern, L.~J.~Dixon, D.~C.~Dunbar and D.~A.~Kosower,
  ``One loop n point gauge theory amplitudes, unitarity and collinear limits,''
  Nucl.\ Phys.\  B {\bf 425}, 217 (1994)
  [arXiv:hep-ph/9403226].

\bibitem{unexp}
  Z.~Bern, J.~J.~Carrasco, D.~Forde, H.~Ita and H.~Johansson,
  ``Unexpected Cancellations in Gravity Theories,''
  Phys.\ Rev.\  D {\bf 77}, 025010 (2008)
  [arXiv:0707.1035 [hep-th]].

\bibitem{sok2}
  J.~M.~Drummond, J.~Henn, G.~P.~Korchemsky and E.~Sokatchev,
  ``Generalized unitarity for N=4 super-amplitudes,''
  arXiv:0808.0491 [hep-th].

 \bibitem{qmc}
  A.~Brandhuber, P.~Heslop and G.~Travaglini,
  ``A note on dual superconformal symmetry of the N=4 super Yang-Mills
  S-matrix,''
  arXiv:0807.4097 [hep-th].


\bibitem{gris}
  M.~T.~Grisaru and H.~N.~Pendleton,
  ``Some Properties Of Scattering Amplitudes In Supersymmetric Theories,''
  Nucl.\ Phys.\  B {\bf 124}, 81 (1977).


\bibitem{bernpc} Z.~Bern, private communication.

\bibitem{dhks}
  J.~M.~Drummond, J.~Henn, G.~P.~Korchemsky and E.~Sokatchev,
  ``Dual superconformal symmetry of scattering amplitudes in N=4
  super-Yang-Mills theory,''
  arXiv:0807.1095 [hep-th].

\bibitem{dks}
  J.~M.~Drummond, G.~P.~Korchemsky and E.~Sokatchev,
  ``Conformal properties of four-gluon planar amplitudes and Wilson loops,''
  Nucl.\ Phys.\  B {\bf 795}, 385 (2008)
  [arXiv:0707.0243 [hep-th]].

\bibitem{alday}
  L.~F.~Alday and J.~M.~Maldacena,
  ``Gluon scattering amplitudes at strong coupling,''
  JHEP {\bf 0706}, 064 (2007)
  [arXiv:0705.0303 [hep-th]].

\bibitem{nima2}
  N.~Arkani-Hamed, F.~Cachazo and J.~Kaplan,
  ``What is the Simplest Quantum Field Theory?,''
  arXiv:0808.1446 [hep-th].









\end{thebibliography}
\end{document}